\documentclass[%
a4paper,							
11pt,								
bibliography=totoc,						
abstracton,					
]
{scrartcl}

\usepackage[a4paper,left=3.0cm,right=2.5cm, top=2.5cm, bottom=3.0cm]{geometry}
\usepackage[headsepline=.4pt,footsepline=.4pt,automark,autooneside=false,]{scrlayer-scrpage}
\clearscrheadfoot 
\automark[subsection]{section} 
\automark*[subsubsection]{subsection}
\ihead{\scriptsize\rightmark} 
\usepackage[ngerman,english]{babel}
\usepackage[T1]{fontenc}
\usepackage[utf8]{inputenc}
\usepackage{lmodern} 
\usepackage[onehalfspacing]{setspace} 

\usepackage{xspace}
\usepackage{calc}

\usepackage{amsmath}
\usepackage{amssymb}
\usepackage{amsthm,thmtools}

\usepackage{mathtools}
\usepackage{bbm}
\usepackage{bm}
\begingroup
    \makeatletter
    \@for\theoremstyle:=definition,remark,plain\do{%
        \expandafter\g@addto@macro\csname th@\theoremstyle\endcsname{%
            \addtolength\thm@preskip\parskip
            }%
        }
\endgroup

\usepackage{chngcntr}
\counterwithin*{equation}{section}

\theoremstyle{plain}
\newtheorem{theorem}{Theorem}[section]

\theoremstyle{definition}

\theoremstyle{remark}
\newtheorem{remark}[theorem]{Remark}
\theoremstyle{plain}

\theoremstyle{plain}

\theoremstyle{remark}
\newtheorem{example}[theorem]{Example}
\usepackage{tabularx}
\usepackage{booktabs}
\usepackage{multirow}
\usepackage{multicol}
\usepackage{pdflscape}

\usepackage{diagbox}

\usepackage[flushleft]{paralist}
\setdefaultenum{(i)}{(a)}{(1)}{(aa)}

\usepackage{color}
\usepackage{graphicx}
\usepackage{tikz}
\usetikzlibrary{calc,spy,backgrounds,arrows,intersections}
\usepackage[final]{showkeys}

 \usepackage[normalem]{ulem}
\usepackage{fancyvrb}

\usepackage{float}
\usepackage[section]{placeins}
\usepackage[format=plain]{caption}
\usepackage{algorithm2e}
\usepackage{csquotes}
\usepackage[
	backend=bibtex,
	style=bwl-FU,
	natbib=true,
	sortcites=true,
	block=space
]{biblatex}
\DeclareFieldFormat{postnote}{\xspace\textit{#1}}
\DeclareFieldFormat{multipostnote}{\xspace\textit{#1}}

\usepackage{titleref}
\usepackage{hyperref}
\usepackage{cleveref}
\hypersetup{
    colorlinks,
		linkcolor={red},
    citecolor={blue},
    urlcolor={blue}
}

\renewcommand{\tilde}[1]{\widetilde{#1}}

\newcommand{\1}{\mathbbm{1}}
\renewcommand{\P}{\mathbb{P}}
\newcommand{\Q}{\mathbb{Q}}
\newcommand{\R}{\mathbb{R}}
\newcommand{\N}{\mathbb{N}}
\newcommand{\norm}[1]{\left\|#1\right\|}
\newcommand{\abs}[1]{\left|#1\right|}

\newcommand{\PD}{\mathrm{PD}}

\newcommand{\evoSys}{U}
\newcommand{\semiGroup}{U}

\newcommand{\generator}{A}
\newcommand{\stateSpace}{S}
\newcommand{\ratingMatrix}{R}

\newcommand{\exposure}{\mathrm{V}}

\newcommand{\lgd}{\mathrm{LGD}}

\newcommand{\CBVA}{\mathrm{BVA}}
\newcommand{\CDVA}{\mathrm{DVA}}
\newcommand{\CCVA}{\mathrm{CVA}}
\newcommand{\textCBVA}{$\mathrm{BVA}$\xspace}
\newcommand{\textCDVA}{$\mathrm{DVA}$\xspace}
\newcommand{\textCCVA}{$\mathrm{CVA}$\xspace}
\newcommand{\textCXVA}{$\mathrm{XVA}$\xspace}
\newcommand{\collateral}{\mathrm{C}}

\newcommand{\ratingProcess}{\mathcal{R}}

\newcommand{\CPU}{Intel(R) Core(TM) i7-8750H CPU @ 2.20\,GHz\xspace}
\newcommand{\RAM}{2x32\,GB (Dual Channel) Samsung SODIMM DDR4 RAM @ 2667 MHz\xspace}

\newcommand{\OS}{Windows 10 Pro\xspace}

\SaveVerb{matlab}=Matlab 2021a=
\newcommand{\matlab}{\protect\UseVerb{matlab}\xspace}
\SaveVerb{ga}=ga=

\SaveVerb{fmincon}=fmincon=
\newcommand{\fmincon}{\protect\UseVerb{fmincon}\xspace}
\newcommand{\matlabGOtoolbox}{(Global) Optimization Toolbox\xspace}

\newcommand{\rootFigFolderFOBB}{Figures}
\newcommand{\figureSuffix}{-eps-converted-to.pdf}

\newcommand{\ctimeFminTotalFOBBvalue}{
4.77}

\newcommand{\ctimeSimPTotalFOBBvalue}{
1.69}

\newcommand{\ctimeFminTotalSUBBvalue}{
 15}

\newcommand{\ctimeSimPTotalSUBBvalue}{
7.66}

\newcommand{\errorsFminFOBBOnevalue}{
1.52e-07}
\newcommand{\errorsAnalyticPMarketFOBBOnevalue}{
2.69e-06}
\newcommand{\errorsAnalyticQPDFOBBOnevalue}{
1.85e-06}
\newcommand{\errorsAnalyticPSimPFOBBOnevalue}{
0.000179}
\newcommand{\errorsAnalyticQSimQFOBBOnevalue}{
0.000332}
\newcommand{\errorsFminFOBBTwovalue}{
8.99e-08}
\newcommand{\errorsAnalyticPMarketFOBBTwovalue}{
2.35e-05}
\newcommand{\errorsAnalyticQPDFOBBTwovalue}{
4.93e-09}
\newcommand{\errorsAnalyticPSimPFOBBTwovalue}{
0.000469}
\newcommand{\errorsAnalyticQSimQFOBBTwovalue}{
0.00077}
\newcommand{\errorsFminFOBBThreevalue}{
6.91e-08}
\newcommand{\errorsAnalyticPMarketFOBBThreevalue}{
0.000101}
\newcommand{\errorsAnalyticQPDFOBBThreevalue}{
1.31e-08}
\newcommand{\errorsAnalyticPSimPFOBBThreevalue}{
0.000632}
\newcommand{\errorsAnalyticQSimQFOBBThreevalue}{
0.00121}
\newcommand{\errorsFminFOBBFourvalue}{
6.37e-08}
\newcommand{\errorsAnalyticPMarketFOBBFourvalue}{
0.000464}
\newcommand{\errorsAnalyticQPDFOBBFourvalue}{
6.51e-09}
\newcommand{\errorsAnalyticPSimPFOBBFourvalue}{
0.000806}
\newcommand{\errorsAnalyticQSimQFOBBFourvalue}{
0.00172}

\newcommand{\figCollateralRTFOBBTwo}{%
	\includegraphics[width=\columnwidth]{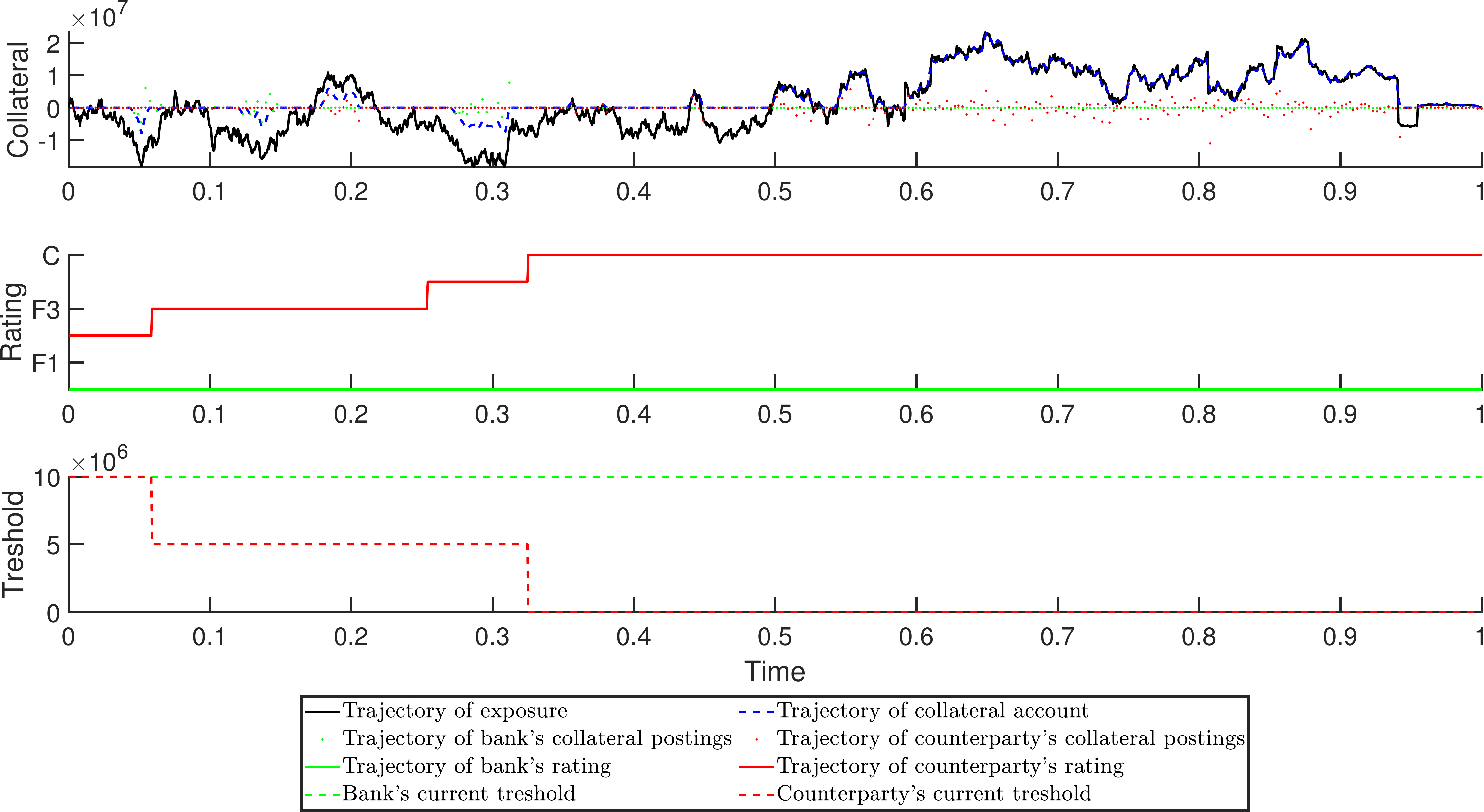}%
}

\newcommand{\figPrePDPFOBBSeven}{%
	\includegraphics[width=\columnwidth]{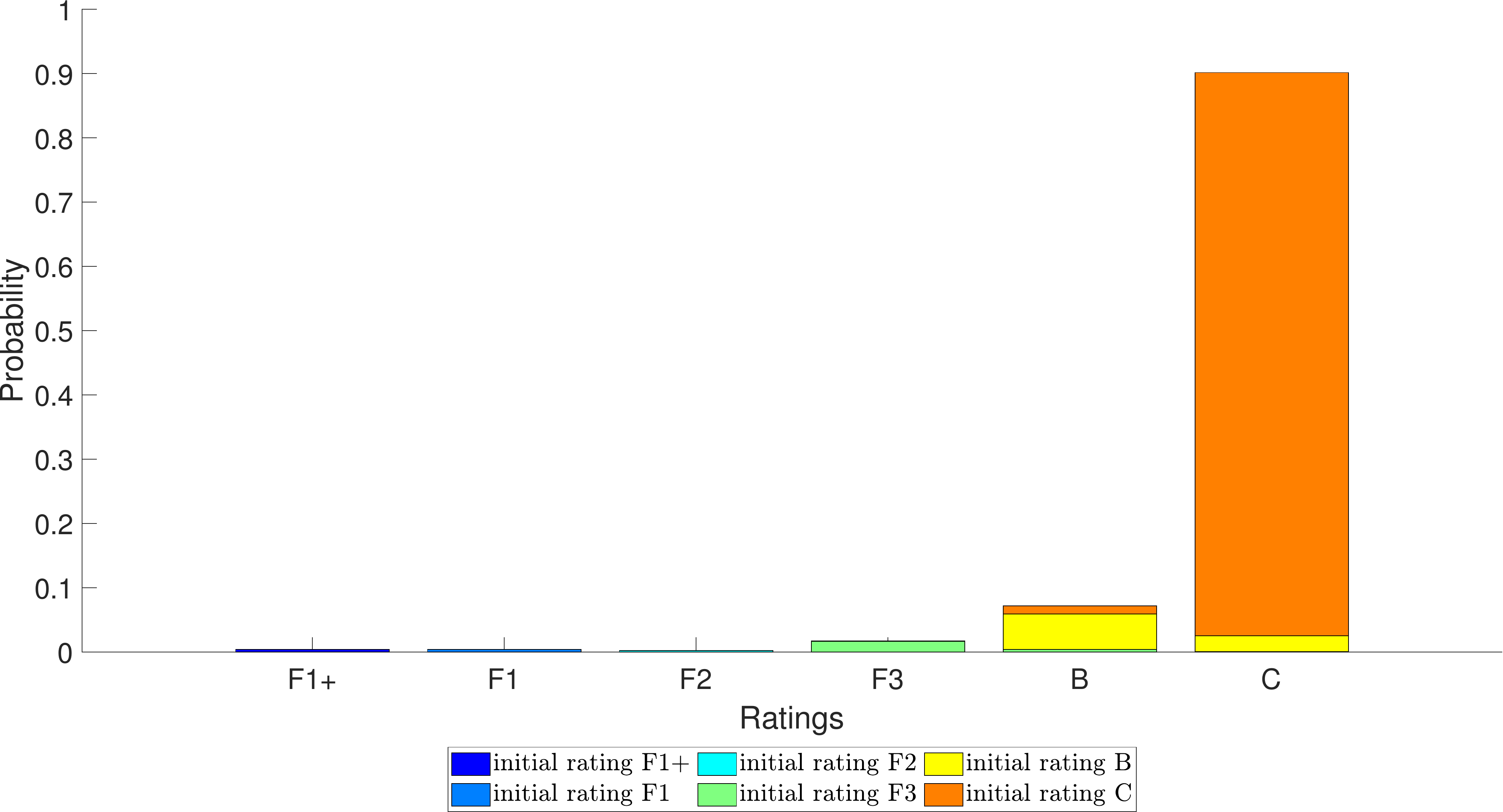}%
}
\newcommand{\figPrePDQFOBBSeven}{%
	\includegraphics[width=\columnwidth]{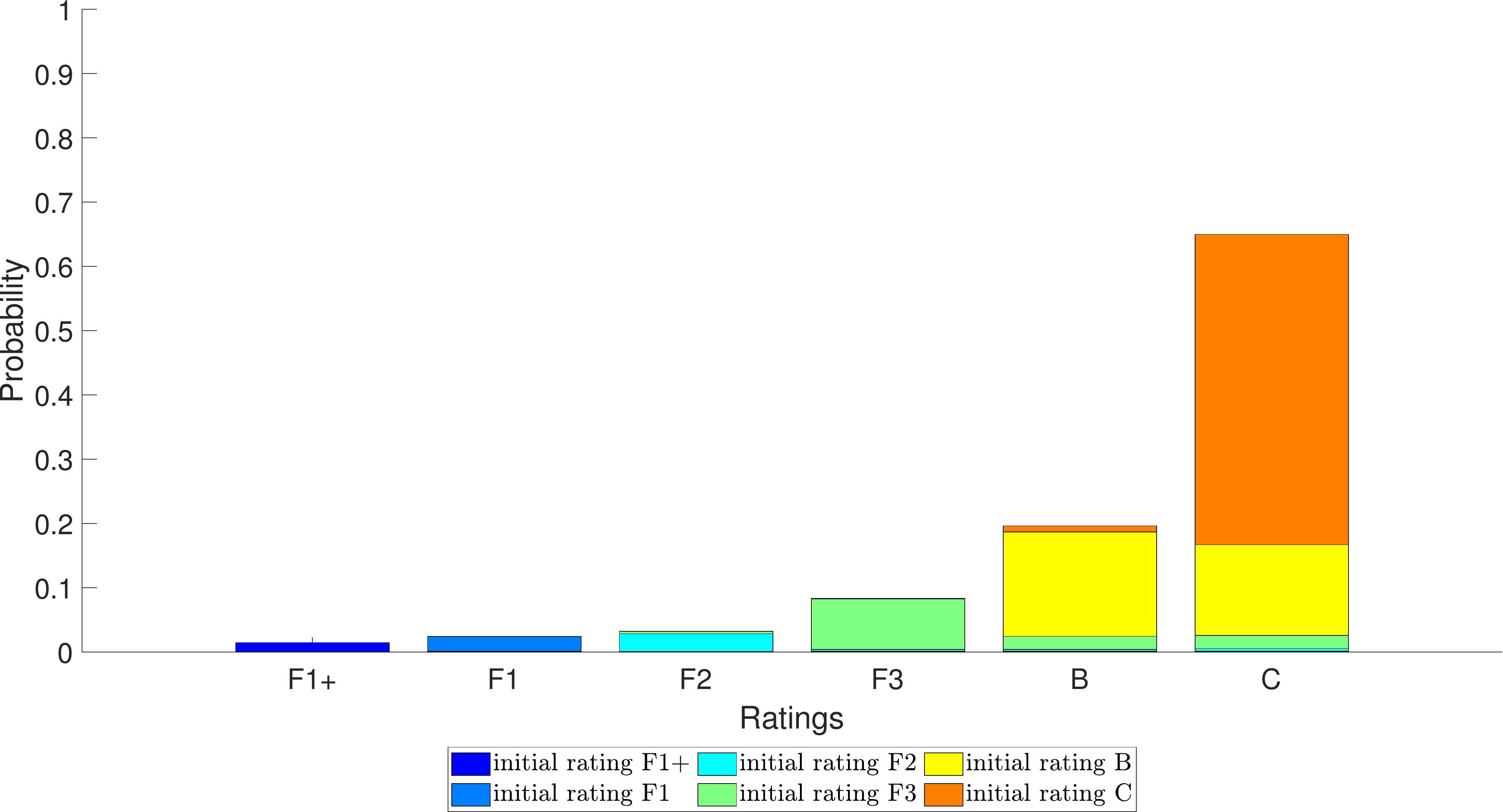}%
}
\newcommand{\figRatingPlotsPFOBBOne}{%
	\includegraphics[width=\columnwidth]{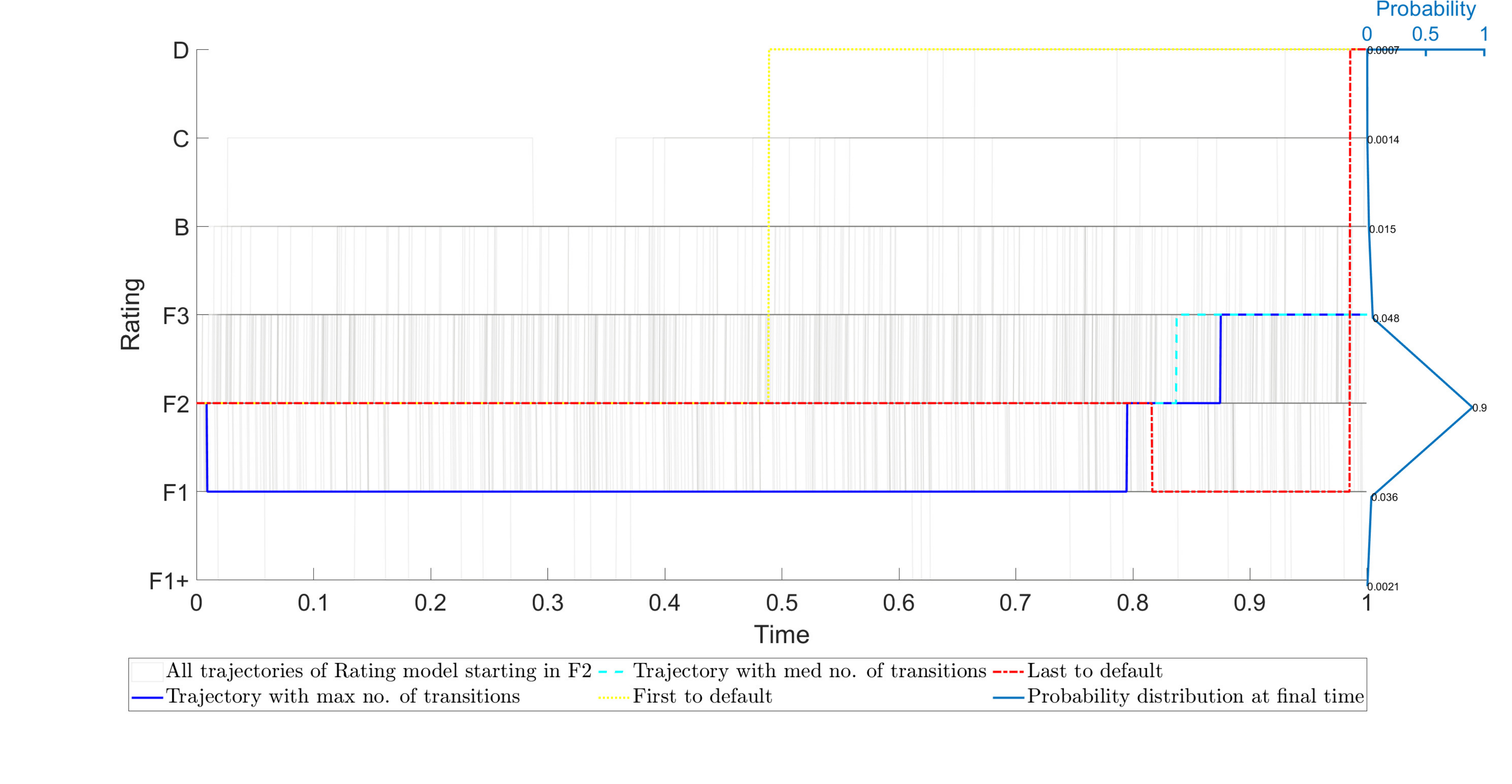}%
}
\newcommand{\figRatingPlotsQFOBBOne}{%
	\includegraphics[width=\columnwidth]{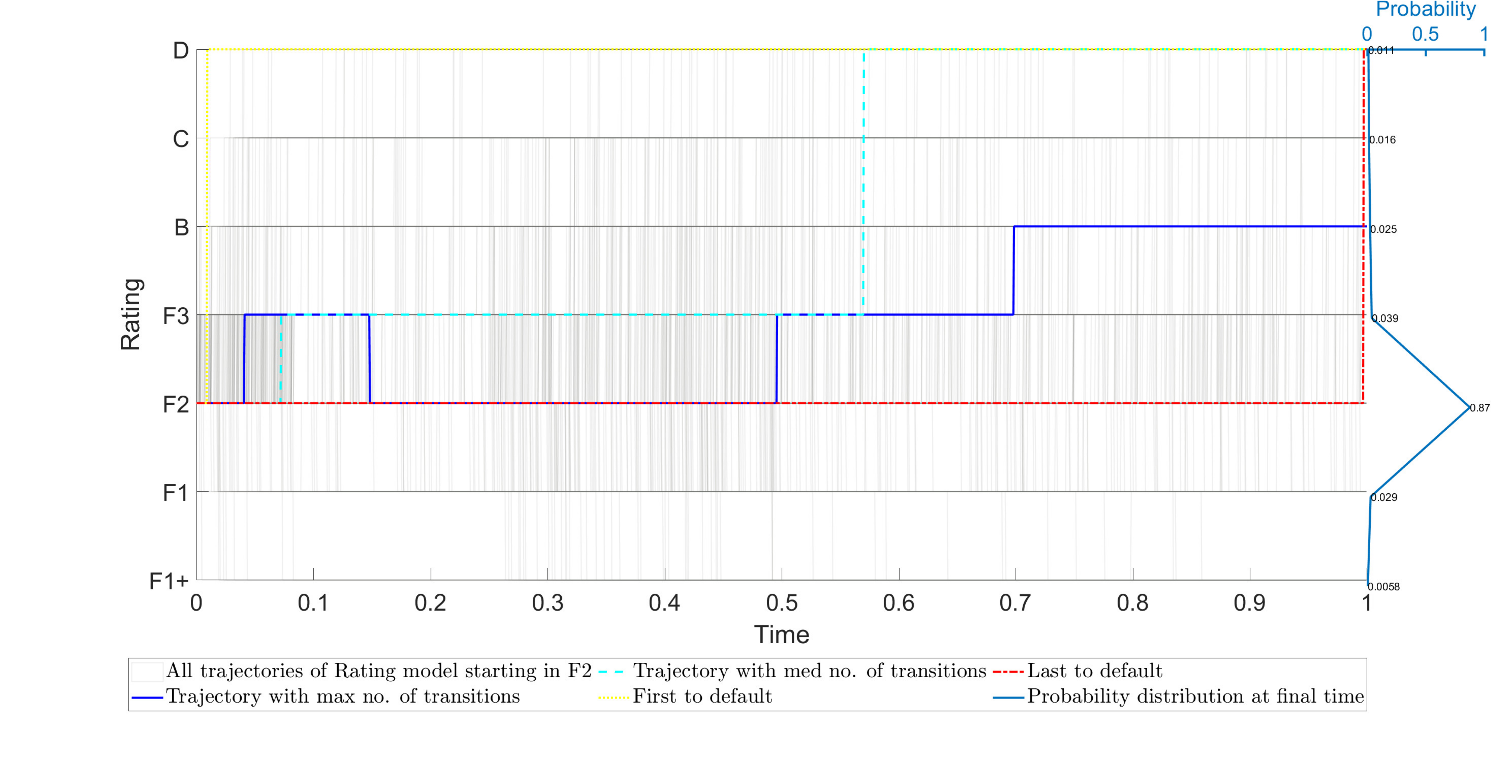}%
}

\title{An introduction to rating triggers for collateral-inclusive XVA in an ICTMC framework}
\author{%
Kevin Kamm%
\thanks{Dipartimento di Matematica, Universit\`a di Bologna, Bologna, Italy. %
\textbf{e-mail}: kevin.kamm@unibo.it}%
}

\begin{document}
\thispagestyle{empty}\pagenumbering{roman}
\maketitle
\renewcommand{\thefootnote}{\Roman{footnote}}
\renewcommand{\thefootnote}{\arabic{footnote}}
\begin{abstract}
In this paper, we model the rating process of an entity as a piecewise homogeneous continuous time Markov chain.
We focus specifically on calibrating the model to both historical data (rating transition matrices) and market data (CDS quotes), relying on a simple change of measure to switch from the historical probability to the risk-neutral one.
We overcome some of the imperfections of the data by proposing a novel calibration procedure, which leads to an improvement of the entire scheme.
We apply our model to compute bilateral credit and debit valuation adjustments of a netting set under a CSA with thresholds depending on ratings of the two parties.
\end{abstract}
\textbf{Keywords:} 
Ratings, Rating Triggers, Default Probability, First to default, Inhomogeneous Markov processes, ICTMC, Exponential Change of Measure, XVA.\\\noindent
\textbf{Acknowledgements:}
This project has received funding from the European Union’s Horizon 2020 research and innovation programme
under the Marie Sklodowska-Curie grant agreement No 813261 and is part of the ABC-EU-XVA
project.

Special thanks goes to the whole team of Banco Santander's XVA desk in Madrid, in particular to my supervisor Luca Caputo,
Michael Belk and Jerome Vincent Maetz for great discussions and help in implementing this approach.
\newpage
\pagestyle{scrheadings}\ihead{\scriptsize\rightmark}\pagenumbering{arabic}
\section{Introduction}\label{sec:introduction}
In this paper, we study bilateral credit and debit valuation adjustments (hereafter referred to as \textCCVA and \textCDVA) of a portfolio of trades between two parties having signed a collateral agreement dependent on ratings. 
Recall that \textCCVA and \textCDVA are adjustment to the fair price of a financial portfolio accounting for the potential loss in case of default of the counterparty and the owner, respectively. These are usually defined as an average of the exposure (positive and negative, respectively) weighted by the probability of default. The two parties usually sign a so-called netting agreement, so as to consider the exposure at portfolio level (as opposed to trade-wise). Attached to the netting agreement, one often has a Credit Support Annex (CSA) by which each of the two parties further agrees to interchange securities (referred to as collateral) to reduce the exposure of the other party. In the case of bankruptcy the collateral account can be used to mitigate the losses of the non-defaulting party, although collateral is often non-segregated and therefore also at risk. Since posting collateral is another expense for an entity, it is desirable to keep the postings as small as possible while simultaneously keeping the losses due to a default event small as well. To this goal, more and more CSAs specify thresholds of permitted unsecured exposure in terms of the credit quality of the parties: the higher the credit quality of a party, the smaller the amount collateral it has to post (and the larger the unsecured exposure of the other party). 

A customary way to measure the credit quality of an entity is to use credit ratings. A high rating means that the entity is very likely to fulfill its financial obligations towards its contracting party, while a low rating associates an increasing risk for meeting the aforementioned obligations. In this line of thought, the default can be viewed as the worst possible rating. CSAs dependent on ratings are said to have rating triggers: a change of rating of one of the parties triggers a change of threshold of that party.

Since the exposure depends on the amount of collateral posted or received, to compute \textCCVA and \textCDVA in presence of a CSA with rating triggers it is necessary to model the rating processes of the contracting parties. As always in modeling, there are trade-offs between describing the real-world behavior and numerical or analytical tractability. We decided in favor of numerical tractability to use Markov processes. The  memoryless property of such processes make them somewhat unrealistic as models for the rating process, since an entity with a history of successive downgrades is more likely to be considered risky than a competitor with long-time constant rating. Another aspect  to take into account is whether the rating process can be assumed to be time homogeneous. It is empirically evident that a time homogeneity assumption is often violated. However, the data published by rating agencies is scarce, for example for short times, e.g. one year, only one, three, six and twelve month rating matrices are usually available. This is precluding the possibility to fit a totally inhomogeneous process to the data in a meaningful way. 
In most of the literature, this motivates the usage of pure time-homogeneous models. However, 
we decided to use piecewise homogeneous continuous time Markov chains (PHCTMC) to model the rating evolutions to capture at least some time-dependent features of the rating evolution. Both assumption are discussed by \cite{Lencastre2014} in greater details.

We calibrate the rating processes simultaneously to historical rating matrices published by rating agencies (such as Moody's, S\&P and Fitch) and default probabilities, derived from Credit Default Swaps (CDS) market quotes.
These are not only two different sources of data inputs but also have a different underlying probability space. Rating transitions are considered under the historical measure, while CDS market quotes are under a risk-neutral probability measure. Hence, it is necessary to have an explicit change of measure formula to ensure a fast calibration procedure.


In the literature, this problem is overcome by extracting the generator of the underlying Markov chain from the rating transition data and calibrating a parametrized change of measure formula to the CDS quotes afterwards. Note, that the risk-neutral measure is an output of the calibration procedure. We will comment on this later on in Remark \ref{rem:errorInMeasure} in more detail.

We found that this split procedure does not perform very well and use instead a penalized calibration procedure allowing for some error with respect to the previously obtained the market generators. This means that the output of the calibration procedure is simultaneously a generator under the historical measure and the parameters for the change of measure. This type of calibration procedure seems to be novel compared to the results in the literature and performs very well.

To assess the performance of this model we will compare three different collateral agreements: an agreement with perfect collateralization, no collateralization and with rating triggers. As expected the model with rating triggers will reduce the losses due to default events but its performance is dependent on the immediate jumps from high ratings to default. We will call the distribution of jumps prior to default the pre-default distribution and study it in the end of the paper.

\subsection{Review of the literature and comparison}\label{sec:review}
Rating models have been studied for several years. As described in \cite[\ p.~76 Section 4.12.1 Standing Assumptions]{Bielecki2003} one can identify two different approaches to rating modeling.

On the one hand, one can model ratings in a HJM-framework, independently proposed by 
\cite{Bielecki2000} and \cite{Schonbucher2003}, 
and, on the other hand, there are intensity-based models, introduced by the pioneering work of \cite{Jarrow1997}. 
As this paper also belongs to the intensity-based models, we will briefly summarize the available literature on the intensity-based approach, focusing particularly on those results which provide context to our results. 

The characteristic of this approach is that the historical and risk-neutral measure are given and the rating model itself is defined under a third measure, which is calibrated to the risk-neutral measure. Usually, Markov processes are used in this framework, in particular CTMCs, whose generator takes the role of the intensities known from default modeling. However, contrary to default modeling, one is not only interested in the first jump time of the process but in all transitions over time from one state to the other.

In \cite{Jarrow1997} the authors propose a continuous-time Markov chain model for the rating process. To tackle the problem of historical versus risk-neutral data they start with a CTMC under the historical measure and assume that there exists a risk-neutral generator given by $A_t = \mathrm{diag}\left(\mu_1(t),\dots,\mu_{K-1}(t),1\right) A$, where
$A$ denotes the generator under the historical measure and $\mu_i(t)$ are positive integrable functions. In the main application the $\mu_i$ are assumed to be constants which amounts to assuming that the rating process is a  time-homogeneous CTMC. The coefficients $\mu_i$, which are calibrated to credit risky bonds, can be thought as risk-premia relating the historical measure to the risk-neutral measure, although the underlying change of measure is not described explicitly in the paper (a rigorous proof that such a change of measure exists can be found in \cite[\ p.~12 Example 2.9.]{Bielecki2009}).

In \cite{Bielecki2012} the authors are interested in bilateral CVA under rating triggers, as well, but are focusing on so-called \emph{Close-outs}, namely clauses dictating the termination of the portfolio whenever a given credit rating is reached by one of the parties. To model the rating evolution they use a Markov Copulae for multivariate time-homogeneous Markov chains to include the possibility of calibrating the rating processes of different sectors in a consistent way. We will discuss this issue further in Section \ref{sec:ratingTriggers}. For the necessary change of measure they apply, in contrast to \cite{Jarrow1997}, an exponential change of measure technique proposed by \cite{Palmowski2002}, which we will also use in this paper. 

In more recent works of Bielecki et al. (\cite{Bielecki2015a} and \cite{Bielecki2015b}) 
conditional Markov chains and Markov copulae approaches are proposed and analysed in a more general mathematical framework.

A detailed overview of CTMC approaches can be found in \cite{Bielecki2009} in which they also discuss the time-inhomogeneous case briefly. 
Additionally, in \cite[pp.~351\,ff. Chapter 12 Markovian Models for Credit Migrations]{Bielecki2004} is a detailed overview of further models for credit migrations.


We will compare our work mainly to \cite{Jarrow1997} and \cite{Bielecki2012}.
Also we will elaborate how the different change of measure techniques therein relate to each other.

Like \cite{Bielecki2012} we are using CTMCs but for simplicity we are not using a copulae approach, which would be more realistic and is subject to future research. However, we are using a simple inhomogeneous extension instead of time-homogeneous CTMCs. For the calibration they are restricting the space of possible parameters, which will not be necessary in our case and allows for better calibration results.


Thus, the main contribution of this paper is to propose a model to price \textCCVA and \textCDVA in presence of collateralization with rating triggers in the context of piecewise-homogeneous continuous time Markov chains. 
Moreover, we 
propose a simple calibration procedure with major improvements to straightforward techniques. Last but not least, we highlight the importance of studying the pre-default rating distribution in the context of collateralization with rating triggers.

The paper is organized as follows. In Section \ref{sec:changeOfMeasureMain} we introduce the mathematical framework of inhomogeneous continuous-time Markov chains and discuss the change of measure technique. This is followed by a brief discussion about extracting a generator from real data, called \emph{embedding problem}, in Section \ref{sec:embedding}. Then, we start with the numerical tests in Section \ref{sec:numerics}. We will first explain the data we use and its issues in Section \ref{sec:market_data} and explain how to calibrate the PHCTMC model to the data in Section \ref{sec:calibration}. Afterwards, we will use the Gillespie Stochastic Simulation Algorithm to simulate our PHCTMC under both measures and use the simulated processes as our rating processes in the context of rating triggers in Section \ref{sec:ratingTriggers}. In Section \ref{sec:CXVA} we will first calculate \textCCVA, \textCDVA and \textCBVA and dicuss the pre-default rating distribution and its impact on \textCXVA in Section \ref{sec:preDefault}.
Last but not least, we will conclude the paper in Section \ref{sec:conclusion} and discuss future extensions and research.

\section{ICTMCs and an exponential change of measure formula}\label{sec:changeOfMeasureMain}

We begin this section by describing the features making CTMC a good choice for a rating model. First, it is a natural choice to consider a discrete state-space consisting of all ratings and a continuous-time framework. Moreover, rating transition probabilities have to be time and state-dependent to reflect the different intensities associated to the different rating-changes over time. About this dependency, two aspects must be considered (see \cite[Section 8.2]{Schonbucher2003} for more details): time homogeneity and the Markov property. Both assumptions are not completely realistic although time homogeneity can be seen as a stronger restriction than the Markovianity assumption. 

As far as calibration is concerned, it is desirable for the model to be consistent with all the available data, namely both the historical rating transition matrices and the market quoted credit spreads (either coming from bonds or from credit default swaps). For this to be feasible, one needs a change of measure formula, preferably preserving some of the feature of the model (for instance, Markovianity).

Finally, for numerical and/or analytical tractability of the model is certainly a fundamental need. This led us to consider Markov processes for the rating process. CTMCs are then a natural choice in view of the above (continuous-time, a discrete state-space and Markovianity). As far as time dependence is concerned, we propose a simple non-homogeneous CTMC model, namely a piecewise-homogeneous CTMC (PHCTMC). 

Let us briefly recall the relevant terminology of ICTMCs (cf. \cite[pp.~326\,ff. Chapter 11 Markov Chains]{Bielecki2004}). An ICTMC is a Markov process $\left(X_t\right)_{t\in [0,T]}$ with a discrete state space $\stateSpace$. We will only consider the case of a finite state space, i.e. $\stateSpace=\left\{1,\dots,K\right\}$. The functions $p_{ij}(s,t)\coloneqq \P\left(\left.X_t=j\right|X_s=i\right)$ are called \emph{transition functions} and their time-derivative from above is called \emph{generator} of the ICTMC, i.e. $\generator_{ij}(t) \coloneqq \lim_{h\downarrow 0} \frac{p_{ij}(t,t+h)-\delta_{ij}}{h}$, $i,j=1,\dots,K$ and $\delta_{ij}$ is the Dirac-delta. We have the following immediate properties of generator: $\generator_{ij}(t)\geq 0$ for all $i\neq j$ and 
$\generator_{ii}(t)=-\sum_{j\neq i}^{}{\generator_{ij}(t)}$.

For the change of measure we have the following result from \cite[pp.~13--15 Section 3 The Equivalent Martingale Measure]{Ding2021} or \cite[pp.~334\,ff. Chapter 11.2.5 Change of Probability Measure]{Bielecki2004}:
\begin{theorem}\label{thm:main}%
	Let $\left(X_t\right)_{0\leq t \leq T}$ be an ICTMC on the probability space 
	$\left(\Omega,\mathcal{F},\left(\mathcal{F}_t\right),\P\right)$ taking values in $\stateSpace \coloneqq \left\{1,\dots, K\right\}$ with generator $\generator^{\P}_t \coloneqq \left(\generator_{i,j}^{\P}(t)\right)_{i,j=1,\dots,K}$. Define 
	\begin{align*}
		L_t^\kappa \coloneqq
		\exp\left(-M_t\right)
		\prod_{0<u\leq t}^{}{
			\left(
				1+
				\sum_{i,j=1}^{K}{
					\kappa_{ij}(u)
					\left(
						H_{ij}(u) - H_{ij}(u-)
					\right)
				}
			\right)
		},
	\end{align*}
	where
	\begin{compactenum}
		\item the stochastic processes $\kappa_{i,j}(t)$ are bounded, real-valued and $\mathbb{F}$-predictable, such that $\kappa_{ij}(t)>-1$ and $\kappa_{ii}(t)=0$;
		\item the number of jumps from rating $i$ to $j$ are $H_{ij}(t)\coloneqq \sum_{0\leq u \leq t}^{}{\1_{X_{u-}=i}\1_{X_{u}=j}}$;
		\item 
			$
				M_t\coloneqq 
				\int_{0}^{t}{
					\sum_{i,j=1}^{K}{
						\kappa_{ij}(u) \generator^{\P}_{i,j}(u)
						\1_{X_u=i}
					}
					du
				}.
			$
	\end{compactenum}
	Then $L_t^\kappa$ is a strictly positive martingale under $\P$ satisfying $\mathbb{E}^{\P}\left[L_T\right]=1$ and the equivalent probability measure $\Q^\kappa$ given by
	\begin{align*}
		\left. 
			\frac{d\Q^\kappa}{d\P}
		\right|_{\mathcal{F}_t} = L^\kappa_t
	\end{align*}
	$\P$-almost surely, is well-defined. 
	
	Furthermore, $X_t$ is an ICTMC under $\Q^\kappa$ as well with generator
	\begin{align*}
		\generator_{i,i}^\kappa(t) &= -\sum_{i\neq j}^{}{\generator_{i,j}^\kappa(t)}\\
		\generator_{i,j}^\kappa(t) &= \left(1+\kappa_{i,j}(t)\right)\generator_{i,j}^{\P}(t).
	\end{align*}
\end{theorem}
We can see that for any admissible family $\kappa\coloneqq \left(\kappa_{ij}\right)_{i,j=1,\dots,K}$ we get a valid change of measure and know the generator of the ICTMC after the change of measure as well.

As mentioned in the introduction, there are two major examples of this change of measure in the rating community.
\begin{example}\label{exm:changeOfMeasure}%
	On the one hand, if we set $\kappa_{ij}(t)\coloneqq h_i(t) -1$ for strictly positive deterministic functions $h_i(t)>0$, $i\neq j$, we recover the change of measure by \cite{Jarrow1997},
	which we will call JLT-change of measure from now on.
	
	On the other hand, if we set $\kappa_{ij}(t) \coloneqq \frac{h_i(t)}{h_j(t)} -1$
	 we get a time-inhomogeneous version
	of the exponential change of measure technique by \cite{Palmowski2002}, which we will
	call exponential change of measure throughout the entire paper.
	
	In both cases, we will denote the resulting measure by $\Q^h$ and the generators under
	this measure by $\generator^h_t$.
\end{example}
The reason for choosing only $K$ different functions $h_i$ is due to the fact that after the change of measure we know only data with respect to the default column, and there seems to be no meaningful way to use an entire matrix $\kappa_{ij}$.

With this change of measure formula, we can calibrate the model (see Section \ref{sec:calibration}) under the historical measure $\P$ and risk-neutral measure $\Q$ by finding the appropriate functions $h$, such that $\Q^h$ and $\Q$ are close with respect to the default probabilities. 

But before we can start with the calibration procedure we need to discuss how one extracts the generator $\generator^{\P}$ from the historical data and associated problems, which is the subject of the following subsection.
\subsection{Embedding problem}\label{sec:embedding}
Another important aspect, which has to be studied for rating models in the context of ICTMCs, is how to extract the generator of an ICTMC from the historical data, because we will need it for the change of measure formula and the simulation of the process. 

In the homogeneous case, the question boils down to asking if the given transition matrix has a matrix logarithm, such that all off-diagonal entries are positive and all row sums are zero.

In general, this is still an open problem and in the literature referred to as \emph{embedding problem}, which is deeply connected to root finding techniques.
The paper by \cite{Lencastre2016} provides a nice overview of the known conditions on the existence of a generator for both the homogeneous and inhomogeneous case.

As aforementioned, in general the embedding problem is not fully studied at this point in time and has only been solved for $2\times 2$ and $3\times 3$ matrices in general. A condition for $4\times 4$ can be found in \cite{Casanellas2021} with a good overview of the existing literature.

For our approach the paper by \cite{Israel2001} is of particular interest, since it deals with the embedding problem in a rating context and provides us with a numerical justification to approximate a generator from given historical data. They also note, that in general the data does not yield a valid generator, therefore an adjustment to either the rating matrices or the extracted generator is necessary. 
Another numerical approach to approximate a generator from given historical data can be found in \cite{Kreinin2001}. They use a best approximation technique in a suitable space for transition matrices.

In Section \ref{sec:calibration} we will apply the simple justification by \cite{Israel2001} and mitigate the effects of this approximation in our calibration procedure. We use the simple approach because this part is not our main focus in this paper. However, we would like to caution the reader that this choice might have a significant impact on the outcome. Further studies would be needed to estimate the sensitivity of this choice but for this more data is required. We will briefly discuss this issue in Section \ref{sec:market_data}: note that the published data does not even yield directly a valid transition matrix due to \emph{withdrawals}, namely those cases where companies stop being rated (for different reasons) during the period of calculation of a rating matrix. 

\section{Numerical implementation of ICTMCs for rating triggers}\label{sec:numerics}
After establishing the change of measure in Theorem \ref{thm:main} for the general case of ICTMCs, we will now consider piecewise homogeneous CTMCs (PHCTMC) $X_\cdot$ in this section. To be more precise, let $T_0\in [0,T]$ be the initial time and $T_k$, $k=1,\dots,n$ be the points in time when historical rating matrices---denoted by $R^M_t$---are available in an increasing order with $T_n=T$, then $X_\cdot$ is assumed to
be homogeneous on each $[T_k,T_{k+1})$ $k=0,\dots,n-1$.

We used for the calculations \matlab with the \matlabGOtoolbox
running on \OS, on a machine with the following specifications: processor
\CPU and \RAM.

The section is organized as follows: First, we have a look at the available historical data and the problems intertwined with the data in Section \ref{sec:market_data}. Then, we will calibrate a PHCTMC to the market data in Section \ref{sec:calibration} and 
we compare the exponential change of measure to the JLT change of measure in Section \ref{sec:comparisonJLT}.
 Last but not least, we simulate the trajectories of the PHCTMC using Gillespie's stochastic simulation algorithm in Section \ref{sec:simulation}.
\subsection{Historical and Market Data}\label{sec:market_data}
Appendix \ref{sec:marketData} contanins the collection of the rating matrices under the historical measure and default probabilities under the risk neutral measure, which we use in this paper.

The result presented in the paper are obtained using the historical data published by Fitch Ratings from 2014 containing one, three, six and twelve month transition matrices. We verified that our method also works with the matrices published by S\&P from 2020 but we decided not to include the results, since they mirror the ones obtained from the first data set. Note that, regardless of the source of data, the rows of historical transition matrices do not in general sum up to one due to companies, which stop being rated during the matrix calculation period (see for instance 
Table \ref{tab:ratingMatrixMarketFOBBFour}).  Unfortunately, it is not known publicly when and with what rating a company drops out of the rating procedure, making it very difficult to reconstruct the data. We will come back to this problem in Remark \ref{rem:longTerm}.

The default probabilities for the different initial ratings are obtained from market CDS spreads published at the beginning of 2022. Specifically we considered the spreads for the the financial sector, without distinguishing by geography. To derive default probability from spreads we used a standard bootstrapping procedure. 

\subsection{Calibration and rating matrix adjustments}\label{sec:calibration}

The calibration procedure has the following three steps:
First, we need to adjust the historical data to remove the withdrawal component, then we have to extract an approximation of the generator and finally calibrate the ICTMC to the previously obtained generators with the exponential change of measure technique.

Let us first of all fix some notations throughout this section:
As mentioned above, $R^M_t$ will denote the historical rating matrix at time $t$ with possible row-sum less than one,
$R^A_t$ will be the adjusted rating matrix with row-sum equal to one and
$\generator^{M}_t$ will denote the approximation of the generator of $R^A_t$.

\paragraph*{Historical rating matrix adjustment} 
As mentioned in Section \ref{sec:market_data}, this step is performed without further knowledge on the time of withdrawal and rating of the company at the time of  withdrawal. When performing the adjustment, one would like to have some general properties the adjusted matrix should satisfy. However, there seems to be no consensus on what properties a rating matrix should have, e.g. monotone increasing rows till the diagonal element and monotone decreasing afterwards. The only property, which is mentioned in the literature is \cite[\ p.~15 Lemma 2 (Credit Ratings Versus Risk)]{Jarrow1997}, ensuring that lower credit ratings should be riskier. Nevertheless, this does not seem to be enough to describe the usual observed shape of a rating matrix and is up to future research to find a suitable mathematical and economical characterization. We noticed that there are three prominent components: first, the default column and row have different properties compared to the rest of the matrix, second the neighboring region to the default column (usually one or two ratings), which seems to stretch over time, will be influenced by the default column over time and last but not least, the rest of the matrix has usually a skewed Gaussian shape around the diagonal, making it more likely to decrease in rating.

We decided to use the approach described in Algorithm \ref{algo:MarketDataAdjustment} for the initial data reconstruction. The effects of this adjustment will be mitigated in the calibration step but this choice might have a significant impact on the outcome of the entire model. First, we iterate over the rows of the input historical rating matrix $R^M_t$. Then we calculate the withdrawal. If the withdrawal is greater than zero, we first decide, that a transition probability should never be zero and set it to a small number. Afterwards we calculate the weights of each rating in the current row by dividing each entry by the total mass of the row. Now, we calculate the adjustment for each entry by multiplying the weights with the withdrawal. The adjusted rating matrix will then be the sum of the adjustment and the current row of the historical rating matrix. If the withdrawal was zero, we set the current row of the adjusted rating matrix to the row of the historical rating matrix.  

\begin{algorithm}[H]
\caption{Adjustment of the historical rating matrices.}
\SetKwInOut{Input}{Input}\SetKwInOut{Output}{Output}
\SetKwData{WD}{wd}\SetKwData{y}{y}\SetKwData{b}{b}
\SetKwData{K}{K}
	\Input{$R_t^M \in \R^{K,K}$ with row-sums less or equal to $1$}
	\Output{$R_t^A \in \R^{K,K}$ with row-sums equal to $1$}
	\BlankLine
	\For{$i\leftarrow 1$ \KwTo \K}{
	\WD$\leftarrow \sum_{j=1}^{\K}{\left(R_t^M\right)_{ij}}$ \;
	\eIf{\WD$ > 0$}{
		\y$\leftarrow \left(R_t^M\right)_{i,j=1,\dots,\K}$\;
		\y$\left(\y==0\right)\leftarrow 1e^{-10}$\;
		\b$\leftarrow \frac{\y}{\sum_{j=1}^{K}{\y_j}}\cdot \left(1-\WD\right)$\;
		$\left(R_t^A\right)_{i,j=1,\dots,\K}\leftarrow \left(R_t^M\right)_{i,j=1,\dots,\K} + \b$
	 }{
		$\left(R_t^A\right)_{i,j=1,\dots,\K}\leftarrow \left(R_t^M\right)_{i,j=1,\dots,\K}$
	 }
	}
	\label{algo:MarketDataAdjustment}
\end{algorithm}

\paragraph*{Market generator extraction}
As discussed in Section \ref{sec:embedding} not all given datasets yield a valid generator. Therefore, an approximation is needed, which we perform in two steps and is described in 
Algorithm \ref{algo:generatorApproximation}. First, we apply the matrix logarithm (cf. \cite{Higham2012} and \cite{Higham2013})
to the adjusted rating matrix. Then, we set the diagonal and any negative entries of the matrix logarithm to zero. For a justification of setting the negative entries to zero we refer to \cite[\ p.~6 Section 3 The Non-Negativity Condition]{Israel2001}.
Now, we sum up the rows and set the new diagonal to the negative value of the sums.

A more sophisticated approach is presented in \cite{Kreinin2001} by using a best-approximation approach in a suitable space.

\begin{algorithm}[H]
\caption{Approximation of the historical generators.}
\SetKwFunction{Logm}{logm}
\SetKwFunction{Abs}{abs}
\SetKwInOut{Input}{Input}\SetKwInOut{Output}{Output}
\SetKwData{A}{A}
\SetKwData{K}{K}
	\Input{$R_t^A \in \R^{K,K}$}
	\Output{$\generator_t^M \in \R^{K,K}$ approximated generator of $R_t^A$}
	\BlankLine
	\A$\leftarrow \Logm \left(R_t^A\right)$\;
	$\A\left(\A<0\right)\leftarrow 0$\;
	\For{$i\leftarrow 1$ \KwTo \K}{
		$\left(\A\right)_{ii}\leftarrow -\sum_{j\neq i}^{}{\left(\A\right)_{ij}}$\;
	}
	$\generator_t^M \leftarrow \A$\;
	\label{algo:generatorApproximation}
\end{algorithm}

\paragraph*{Calibrating the PHCTMC}

As aforementioned, we need to calibrate the rating model to the historical data and market data with their respective measures. On the $\P$-side we have the rating matrices $R^M_t\in \R^{K,K}$, discussed above, and on the $\Q$-side we have the default probabilities $\PD(t)\in \R^{K,1}$ for each initial rating.
	
	Let us start under the historical measure $\P$.
	
	As aforementioned, let $T_0\in [0,T]$ be the initial point and $T_k$, $k=1,\dots,n$ be
	the points in time, when rating matrices are available in an increasing order with
	$T_n=T$, then $X_\cdot$ is assumed to
	be homogeneous on each $[T_k,T_{k+1})$ $k=0,\dots,n-1$.
	
	Now, by the Chapman-Kolmogorov equation we get
	\begin{align}
		\evoSys^{\P}_{T_0,T}=
		\evoSys^{\P}_{T_0,T_1}\cdot\evoSys^{\P}_{T_1,T_2}\cdots \evoSys^{\P}_{T_{n-1},T_n}=
		\prod_{k=1}^{n}{\evoSys^{\P}_{T_{k-1},T_{k}}}.
		\label{eq:piecewiseHomogeneousChapman}
	\end{align}
	By homogeneity on each sub-interval we know that the evolution system will reduce to
	a semigroup and its generator will be time-constant with explicit formula
	\begin{align}
		\evoSys^{\P}_{T_{k-1},t}=
		\semiGroup^{\P}_{t-T_{k-1}}=
		\exp\left(
			\generator_k^{\P} \left(t-T_{k-1}\right)
		\right), \quad t\in [T_{k-1},T_{k}).
		\label{eq:piecewiseHomogeneousGenerator}
	\end{align}
	Hence, to extract these generators from the historical data, we denote by 
	$\ratingMatrix^M_{k}$ the rating 
	matrices at times $T_{k}$, $k=1,\dots,n$ and solve
	\begin{align*}
		\evoSys^{\P}_{T_0,T_{k}}\overset{!}{=}
		\ratingMatrix^M_k,
	\end{align*}
	which is by \eqref{eq:piecewiseHomogeneousGenerator} under the assumption that 
	$\evoSys^{\P}_{T_0,T_{k-1}}$ is invertible and $\left(\evoSys^{\P}_{T_0,T_{k-1}}\right)^{-1}\cdot \ratingMatrix^M_k$ has a matrix logarithm equivalent to
	\begin{align*}
		\generator_k^{\P} =
		\frac{
			\log\left(
				\left(\evoSys^{\P}_{T_0,T_{k-1}}\right)^{-1}\cdot \ratingMatrix^M_k
			\right)
		}{
			T_k-T_{k-1}
		},\qquad 
		\evoSys^{\P}_{T_0,T_{k-1}}=
		\prod_{l=1}^{k-1}{\evoSys^{\P}_{T_{l-1},T_{l}}}.
	\end{align*}
	This yields an iterative scheme on the $\P$-side and we will use the same procedure described in Algorithm \ref{algo:generatorApproximation} to retrieve valid generators at each time.
	
	On the risk-neutral side we proceed similarly. The generator on the $\Q$-side is assumed to have
	the same properties as the one on the $\P$-side, hence the change of measure
	formula in Example \ref{exm:changeOfMeasure} will be piecewise constant on each sub-interval, as well, such that we solve a minimization problem at each time $T_k$ by finding the appropriate vector $h_k\in \R^K_{>0}$.
	
	Additionally, let us assume that the default state is absorbing, i.e. we have 
$e_K^T \ratingMatrix^M_t e_i = \delta_{Ki}$, $i=1,\dots,K$, $t\geq 0$, and therefore
$\generator e_i = 0$. Hence, we can choose the last entry in the vector $\left(h_k\right)_K$ freely in each interval and will assume $\left(h_k\right)_K\equiv 1$ without altering the following minimization problem for finding the right change of measure:	
	\begin{align}
		\label{eq:calibration}
		&\min_{\substack{h_k \in \R^{K,1}_{>0}, \left(h_k\right)_K=1,\\ A_k \in \mathcal{A}}}
			\norm{
				\evoSys_{T_{0},T_{k-1}}^{\Q^L}\cdot
				\exp\left(
					A_k^h
					\left(T_k-T_{k-1}\right)
				\right)\cdot e_K-\PD\left(T_k\right)
			}_{\mu_{\Q}}+
			\norm{
					A_k-\generator_k^{\P}
			}_{\mu_{\P}}
			,\\&\notag
			\mathcal{A} \coloneqq
			\left\{
				A \in \R^{K,K}: \text{ for all $i,j=1,\dots,K$ }
											 A_{K,j}=0,\ A_{i,j}\geq 0 \text{ for $i\neq j$ and }
											 A_{i,i}\leq 0
			\right\}.
			\\&\notag
			A_k^h\coloneqq
				\begin{cases}
					A_k \frac{\left(h_k\right)_j}{\left(h_k\right)_i},& i\neq j,\\ 
					-\sum_{k\neq i}^{}{A_k \frac{\left(h_k\right)_j}{\left(h_k\right)_i}},& i=j,\\ 
				\end{cases}
			\\&\notag
			\evoSys_{T_{0},T_{k-1}}^{\Q^L}=
			\prod_{l=1}^{k-1}{\evoSys^{\Q^L}_{T_{l-1},T_{l}}}.
	\end{align}
	The norms $\norm{\cdot}_{\mu_{\Q}},\norm{\cdot}_{\mu_{\P}}$ are weighted norms and defined as follows: Let $X\in \R^{K,K}$, $x\in \R^{K,1}$ and $M^{\P} \in \R^{K,K}_{\geq 0}$, as well as $m^{\Q}\in \R^{K,1}_{\geq 0}$, we set
	\begin{align*}
		\norm{X}_{\mu_{P}}&\coloneqq \norm{M^{\P}\odot X}_{F},\\
		\norm{x}_{\mu_{Q}}&\coloneqq \norm{m^{\Q}\odot x}_{2},
	\end{align*}
	where $\odot$ denotes the elementwise or Hadamard product. The idea of using weighted sums in the calibration procedure is explained in Remark \ref{rem:weightedNorms} in more detail.
	
	
	Moreover, after this piecewise calibration, the whole right-generator of the ICTMC is then defined as $\generator^P_t = 
	\sum_{k=1}^{n}{\generator^P_k \1_{[T_{k-1},T_k)}(t)}$, $P=\P,\Q$ and to evaluate the
	evolution system one has either to use \eqref{eq:piecewiseHomogeneousChapman}
	or solve the Kolmogorov forward equation with the inhomogeneous generator 
	$\generator^P_t$.
	\begin{table}%
		\caption{One year rating matrix under measure $\Q$ after calibration with $M^{\P}\equiv+\infty$ using Fitch's data.}
		\centering
		\begin{tabular}{|c|*{7}{c}|}
		\hline
		\diagbox{From}{To} & F1+ & F1 & F2 & F3 & B & C & D\\ \hline
		F1+ & $0.000$\,\% & $98.047$\,\% & $0.000$\,\% & $0.000$\,\% & $0.838$\,\% & $0.351$\,\% & $0.764$\,\% \\
		F1 & $0.000$\,\% & $97.033$\,\% & $0.000$\,\% & $0.000$\,\% & $1.468$\,\% & $0.696$\,\% & $0.802$\,\% \\
		F2 & $0.000$\,\% & $89.720$\,\% & $0.000$\,\% & $0.000$\,\% & $5.570$\,\% & $2.860$\,\% & $1.850$\,\% \\
		F3 & $0.000$\,\% & $85.545$\,\% & $0.000$\,\% & $0.000$\,\% & $7.724$\,\% & $4.049$\,\% & $2.683$\,\% \\
		B & $0.000$\,\% & $1.868$\,\% & $0.000$\,\% & $0.000$\,\% & $56.863$\,\% & $32.511$\,\% & $8.758$\,\% \\
		C & $0.000$\,\% & $0.044$\,\% & $0.000$\,\% & $0.000$\,\% & $13.455$\,\% & $71.167$\,\% & $15.334$\,\% \\
		D & $0.000$\,\% & $0.000$\,\% & $0.000$\,\% & $0.000$\,\% & $0.000$\,\% & $0.000$\,\% & $100.000$\,\% 
		\\\hline
		\end{tabular}
	\label{tab:FitchWithoutP}
	\end{table}
\begin{remark}\label{rem:weightedNorms}%
	Let us explain why we decided to include the generator under the measure $\P$ in the calibration procedure. First of all, we know that our initial adjustment has been done in an arbitrary way and impacts the final results. In the case, where $M^{\P}\equiv+\infty$, i.e. the calibration is without the generator under $\P$, we get results that look like
	Table \ref{tab:FitchWithoutP}. We see columns consisting of zero entries. This is obviously nothing like we expect a rating matrix to look like. For example, the first column would suggest that it is impossible to transition to or even stay in rating F1+, which is does not make sense, since companies with the best rating usually have the highest probability to stay in their rating. Therefore, we added the generator under $\P$ to the calibration procedure to mitigate the effects of our initial choice of data reconstruction leading to a significant improvement and making the model feasible if one compares Table \ref{tab:FitchWithoutP} to Table \ref{tab:ratingMatrixAnalyticAQFOBBFour}. In 
	Table \ref{tab:ratingMatrixAnalyticAQFOBBFour} we chose $M^{\P}\equiv 1$ and $m^{\Q}\equiv 1$.
	
	As a side-note the choice of restricting the parameter space in the case $M^{\P}\equiv+\infty$ to $h_i\coloneqq e^{\alpha i}$ as in \cite{Bielecki2012} did not help to remove the columns with zero entries and decreased the performance for $M^{\P}\equiv 1$ and $m^{\Q}\equiv 1$.
	
	Having now two different components in the objective function, we added the possibility to add weights corresponding to the user's trust in the data, e.g. they may depend on liquidity and the size of the withdrawal. If the data is assumed to be trustworthy, one can choose the individual weight parameter large, which ensures a better fit in this particular entry.
\end{remark}
\begin{remark}\label{rem:constraints}%
	We tried different versions of \eqref{eq:calibration}. In particular, we 
	adjusted $A_k$ by Algorithm \ref{algo:generatorApproximation} in the objective function but this led to a decrease in accuracy and increase in computational time. Moreover,
	we added the condition that the sum of each row in the generator is supposed to be zero to the constraints, which led to badly conditioned matrices and unpredictable behavior in the calibration algorithm.
	
	Therefore, we decided to leave it unconstrained but keep the bounds on the matrix entries and adjust the outcome of the calibration for the generator under $\P$ by
	Algorithm \ref{algo:generatorApproximation}, which leads to the presented results. 
	To make this more precise, we did not add the constraint to $\mathcal{A}$ that all rows must sum to zero. Therefore, the matrix $A_k$ as an output of the calibration procedure is not necessary a valid generator but we use Algorithm \ref{algo:generatorApproximation} after the calibration procedure to repair it. Let us denote this adjustment $\tilde{A}_k$ for the moment.
	
	The average error of this final adjustment  ($\frac{1}{K^2}\sum_{k,i,j}\abs{\tilde{A}_k-A_k}$) after the calibration procedure was of magnitude $10^{-4}$ for both Fitch and S\&P's datasets, when summing the absolute values of the differences of all entries and dividing by the number of entries, i.e. the error of this adjustment is negligible and makes the calibration algorithm more robust  and justifies why we did not include the ``sum to zero''-constraint in $\mathcal{A}$.
	
	Moreover, we added the constraint that lower credit ratings should be riskier 
	introduced by
	\cite[\ p.~15 Lemma 2 (Credit Ratings Versus Risk)]{Jarrow1997} to the minimization problem as well, but it did not change the results significantly. Conclusively, we decided not to include it in the paper.
	
	In particular, these choices make it possible to implement the calibration procedure as a weighted non-linear least squares problem.
\end{remark}
\begin{remark}\label{rem:errorInMeasure}%
	Notice, that in our calibration procedure, we are in fact introducing an error in measure.
	This is because the default probabilities coming from CDS quotes were derived in a measure $\Q$ and our change of measure is parametrized by parameters $h$.
	
	We are not aware of any literature discussing this issue and its impact is unknown. However this procedure seems to be common practice, since a-priori the Radon-Nikodym derivative in our rating setting is not known.
\end{remark}
We used \matlab's function \fmincon to solve the minimization problem with bounds 
\eqref{eq:calibration} and display the corresponding errors in 
Table \ref{tab:calibrationErrors}. 
For the errors we used the Frobenius norm to compare the resulting rating matrices under $\P$ to the adjusted rating matrices and the Euclidean norm to compare the resulting probabilities of default to the market data under $\Q$. We divided both norms by the number of entries, i.e. $K^2$ for the Frobenius norm and $K$ for the Euclidean norm.
The total computational time for the calibration was\ctimeFminTotalFOBBvalue\ seconds with Fitch's data and\ctimeFminTotalSUBBvalue\ seconds with S\&P's data.
\begin{table}%
\caption{Calibration errors using Fitch's data. First row is the error of \fmincon, second the mean error of the model rating transitions and adjusted market rating transitions and the third row contains the errors under the risk neutral measure of the model and market probabilities of default.}
\centering
\begin{tabular}{|c|*{4}{c}|} 
	\hline
	\diagbox{Error}{Time}&$t=\frac{1}{12}$ & $t=\frac{3}{12}$ & $t=\frac{6}{12}$ & $t=1$\\
	\hline
	\fmincon & \errorsFminFOBBOnevalue & \errorsFminFOBBTwovalue & \errorsFminFOBBThreevalue &\errorsFminFOBBFourvalue\\
	$\frac{1}{K^2}\norm{R_t^{\P} - R_t^A}_{\R^{K,K}}$ & \errorsAnalyticPMarketFOBBOnevalue&\errorsAnalyticPMarketFOBBTwovalue&\errorsAnalyticPMarketFOBBThreevalue&\errorsAnalyticPMarketFOBBFourvalue\\
	$\frac{1}{K}\norm{R_t^{\Q}e_K - \PD(t)}_{\R^{K}}$ &\errorsAnalyticQPDFOBBOnevalue&\errorsAnalyticQPDFOBBTwovalue&\errorsAnalyticQPDFOBBThreevalue&\errorsAnalyticQPDFOBBFourvalue\\
	\hline
\end{tabular}
\label{tab:calibrationErrors}
\end{table}
All results using Fitch's data can be found in Appendix \ref{sec:modelData} starting with the generators after calibration in Appendix \ref{sec:generatorsP}.
\begin{remark}\label{rem:longTerm}%
	We were also interested in the case of long term historical calibrations, i.e. using three or five year rating matrices. However, we noticed that the withdrawal column was the dominant column. Therefore, it is questionable if it makes sense to reconstruct a row of a rating matrix with a withdrawal of e.g. 50 percent without further knowledge.
	
	Another aspect is that the property we hinted at in Section \ref{sec:calibration} of a Gaussian on the diagonal is not justifiable for long-terms. This might be connected to the aging-effects discussed in \cite{Altman1997}: older bonds might have a tendency to be up- or downgraded in a smaller period of time than newer bonds. Whether this effect is taken into account by rating agencies nowadays is not clear and should be addressed in future research.
	
	Additionally, it is worth pointing out that rating agencies could use the so-called Aalen-Johansen estimator (cf. \cite{Lando2002}) instead of a cohort method (sometimes called static pool method) to compute the rating matrices. This would naturally remove the withdrawal rates. 
		Moreover, let us mention that rating agencies are required by ``Rule 17g-7 of the Securities Exchange Act of 1934'' to publish certain rating changes of individual companies. We implemented a cohort method using this data. However, the results were not close to the published rating matrices. Possible reasons for this could be unpublished sensitive rating data and an initial step to remove correlation structures within the data by using additional knowledge about the companies.
		
		We confirmed that S\&P includes such data taking these two points into consideration in their usual subscription for the interested reader.

	Conclusively, for the time being we restrict ourselves to historical transition matrices up to one year published by the rating agencies in this paper.
\end{remark}
\subsection{Comparison to \cite{Jarrow1997} change of measure}\label{sec:comparisonJLT}
To make this comparison, let us first of all recall their change of measure formula. They assume that the risk-neutral generator is given by 
\begin{align*}
	\generator^{\Q} \coloneqq 
	\mathrm{diag}\left(\tilde{h}_{1},\dots,\tilde{h}_{K-1},1\right)
		\generator^{\P},
\end{align*}
where $\tilde{h}_i\geq 0$ for $i=1,\dots,K-1$.

Furthermore, they suppose that the generator under the historical measure can be extracted from the data: here we will use Algorithm \ref{algo:MarketDataAdjustment} and \ref{algo:generatorApproximation} to extract it. 

Contrary to \cite{Jarrow1997} we will compare the methods using CDS quotes instead of risky bonds to calibrate the parameters $\tilde{h}_i$ but this is no major change.

Since this is a homogeneous CTMC, we will first calibrate it to the one year CDS data and compare it to the one year transitions of the PHCTMC.

The calibration works excellent: Since, we use the adjusted rating matrix, $R_t^{\P}$ is the same as $R_t^{A}$ and for the default probabilities we obtain
$\frac{1}{K}\norm{R_t^{\Q}e_K - \PD(t)}_{\R^{K}}=1.79e-10$. It took only $0.2$ seconds to calibrate the values $\tilde{h}$.

However, the rating transitions apart from the default probabilities look rather peculiar under the risk-neutral measure, which are printed in Table \ref{tab:FitchJLT}. In particular, the rating transitions far from the diagonal entries seem too high and consequently the probability to stay in a rating seems too low.

\begin{table}%
	\caption{One year rating matrix under measure $\Q$ after calibration with \cite{Jarrow1997} change of measure using Fitch's data.}
	\centering
	\begin{tabular}{|c|*{7}{c}|}
	\hline
	\diagbox{From}{To} & F1+ & F1 & F2 & F3 & B & C & D\\ \hline
   F1+ & 57.20\,\% & 26.53\,\% & 12.77\,\% & 1.56\,\% & 1.19\,\% & 0.21\,\% & 0.50\,\% \\ 
		F1 & 14.37\,\% & 39.27\,\% & 36.17\,\% & 4.72\,\% & 3.87\,\% & 0.84\,\% & 0.74\,\% \\ 
		F2 &2.47\,\% & 9.88\,\% & 66.04\,\% & 10.00\,\% & 8.37\,\% & 2.10\,\% & 1.11\,\% \\ 
		F3 &1.73\,\% & 5.07\,\% & 39.91\,\% & 18.13\,\% & 23.89\,\% & 7.53\,\% & 3.70\,\% \\ 
		B &0.42\,\% & 1.29\,\% & 12.81\,\% & 13.05\,\% & 42.86\,\% & 20.85\,\% & 8.68\,\% \\ 
		C &0.085\,\% & 0.257\,\% & 3.040\,\% & 4.98\,\% & 26.15\,\% & 50.14\,\% & 15.33\,\% \\ 
		D &0\,\% & 0\,\% & 0\,\% & 0\,\% & 0\,\% & 0\,\% & 100 \,\%
	\\\hline
	\end{tabular}
\label{tab:FitchJLT}
\end{table}

\subsection{Simulation of the ICTMC}\label{sec:simulation}
There are several techniques in the literature concerning the efficient simulation of ICTMCs and its forward equation. For a detailed discussion we refer to 
\cite{Li2012} and \cite{Arns2010} among many others.
However, for our PHCTMC described in the previous Section \ref{sec:calibration} we will use the Gillespie Stochastic Simulation Algorithm (SSA) (cf. \cite{Gillespie2007}), which is also called \emph{Kinetic Monte Carlo (KME)}, on each sub-interval, where the PHCTMC is homogeneous. It turns out that this approach is sufficiently fast, because our state space has few states. The algorithm is displayed in Algorithm \ref{algo:ssa1} and works as follows:

Remember that our PHCTMC is homogeneous on each interval
$[T_{k-1},T_k]$, $k=1,\dots,n$, and now iterate over those intervals, i.e. assume that we are already at $t=T_{k-1}$ with current rating $i$. On each sub-interval we proceed as follows: 
\begin{compactenum}
	\item If $t\leq T_k$ and $\left(\generator_k\right)_{ii} \neq 0$ draw two uniform random numbers $r_1$, $r_2$, otherwise end and set $R^{i_0}_t=i$ on $[t,T_k]$;
	\item Retrieve the exponentially distributed transition waiting time with parameter
		$-\left(\generator_k\right)_{ii}$ as
		\begin{align*}
			\tau = 
			\frac{-\log\left(r_1\right)}{-\left(\generator_k\right)_{ii}}=
			\frac{\log(r_1)}{\left(\generator_k\right)_{ii}}.
		\end{align*}
		If $t+\tau\geq T_k$ set $R^{i_0}=i$ and go to the next interval, starting with step (i), else continue to calculate the next state;
	\item Now, sample from the discrete state transition distribution
		$\left[\frac{\left(\generator_k\right)_{ij}}{-\left(\generator_k\right)_{ii}}\right]_{j\neq i}$. This can be done by choosing the first integer $j$, such that
		$\sum_{k=1,k\neq i}^{j}{\frac{\left(\generator_k\right)_{ij}}{-\left(\generator_k\right)_{ii}}}> r_2$, which is equivalent to
		\begin{align*}
			\min_j
			\sum_{l=1,l\neq i}^{j}{\left(\generator_k\right)_{il}}>-\left(\generator_k\right)_{ii}r_2
		\end{align*}
		Now, go back to (i) with $R^{i_0}_{t+\tau}=j$.
\end{compactenum}

\begin{algorithm}[H]
\caption{Iterative SSA for the PHCMTC starting in $\left\{1,\dots,K\right\}$.}
\SetKwFunction{Log}{log}
\SetKwFunction{Uniform}{Uniform}
\SetKwFunction{cumsum}{cumsum}
\SetKwFunction{findFirst}{findFirst}
\SetKwFunction{update}{update}
\SetKwFunction{Break}{break}
\SetKwInOut{Input}{Input}\SetKwInOut{Output}{Output}
\SetKwData{temp}{temp} 
	\Input{$A \in \R^{n,K,K}$ generator, $i_0\in \{1,\dots,K\}$ initial state, $M\in \N$ number of trajectories}
	\Output{$R_t^{i_0} \in [0,T]\times \R^{1,K,M}$ trajectories of the simulated PHCTMC starting in $i_0$}
	\BlankLine
	\For{$m\leftarrow 1$ \KwTo M}{
		$i\leftarrow i_0$\;
		\For{$k\leftarrow 1$ \KwTo n}{
			\While{$t<T_k$}{
				\eIf{$A_{k,i,i}==0$}{
					\textbackslash\textbackslash Absorbing state\\
					$t\leftarrow T_k$\;
				}{
					\textbackslash\textbackslash Calculate waiting time\\
					$r_1\leftarrow \Uniform\left(0,1\right)$\;
					$\tau\leftarrow\frac{\log\left(r_1\right)}{A_{k,i,i}}$\;
					\textbackslash\textbackslash Update time\\
					\eIf{$t+\tau\geq T_k$}{
						$t\leftarrow T_k$\;
						\Break\;
					}{
						$t\leftarrow t+\tau$\;
					}
					\textbackslash\textbackslash Calculate state transition\;
					$\temp\leftarrow \cumsum\left(\frac{A_{k,i,j=1,\dots,i-1,i+1,\dots,K}}{-A(k,i,i)}\right)$\;
					$r_2\leftarrow \Uniform\left(0,1\right)$\;
					$j\leftarrow \findFirst\left(\temp\geq r_2\right)$\;
					\eIf{$j<i$}{
						$i\leftarrow j$\;
					}{
						$i\leftarrow j+1$\;
					}
				}
				$\update\left(R_t^{i_0}\right)$\;
			}
			$\update\left(R_t^{i_0}\right)$\;
		}
	}
	\label{algo:ssa1}
\end{algorithm}

On the other hand, we could also implement a time-dependent version of the SSA, which is discussed in e.g.
\cite[\ pp.~5\,ff. Section 4 The Stochastic Simulation Algorithm]{Reinhardt2017},
\cite{Purtan2013} and
\cite{Jansen1995} among many others. In particular, we will use the method described
in \cite[\ pp.~7\,ff.]{Prados1997}.
The principal of the algorithm is essentially the same as before, but now the waiting time is a general exponential random variable depending on an integral. This leads to an integral equation $\int_{t}^{t+\tau}{\left(A_{s}\right)_{i,i}ds}=\log\left(r_1\right)$,
where $r_1$ is a uniform random number, in each iteration, which has to be solved for $\tau$ and the performance of the algorithm is highly dependent on the ability of solving this integral equation very fast.
\begin{table}%
\caption{Calibration errors using Fitch's data. First row is the mean error of the model rating transitions and simulated rating transitions under the historical measure. The second row contains the errors under the risk neutral measure of the model and simulated rating transitions.}
\centering
\begin{tabular}{|c|*{4}{c}|} 
	\hline
	\diagbox{Error}{Time}&$t=\frac{1}{12}$ & $t=\frac{3}{12}$ & $t=\frac{6}{12}$ & $t=1$\\
	\hline
	$\frac{1}{K^2}\norm{R_t^{\P} - R_t^{\text{Sim},\P}}_{\R^{K,K}}$ & \errorsAnalyticPSimPFOBBOnevalue&\errorsAnalyticPSimPFOBBTwovalue&\errorsAnalyticPSimPFOBBThreevalue&\errorsAnalyticPSimPFOBBFourvalue\\
	$\frac{1}{K^2}\norm{R_t^{\Q} - R_t^{\text{Sim},\Q}}_{\R^{K,K}}$ &\errorsAnalyticQSimQFOBBOnevalue&\errorsAnalyticQSimQFOBBTwovalue&\errorsAnalyticQSimQFOBBThreevalue&\errorsAnalyticQSimQFOBBFourvalue\\
	\hline
\end{tabular}
\label{tab:calibrationErrors}
\end{table}
It turned out that both algorithms performed the same regarding simulation errors, which are given in the last two rows of Table \ref{tab:calibrationErrors}.
The errors were computed by first calculating the transition matrices from the simulated rating processes $R^{i,P}_t$, $P=\P,\Q$, by counting how many trajectories are at each state and dividing by the total amount of trajectories. The result of this is denoted by 
$R^{\text{Sim},P}_t$ and we used the Frobenius norm divided by the squared number of ratings to evaluate the error.
However, the iterative version was significantly faster than a naive implementation of the general scheme, i.e.
roughly\ctimeSimPTotalFOBBvalue\ seconds using Fitch's data seconds compared 500 seconds. For S\&P's data Algorithm \ref{algo:ssa1} took \ctimeSimPTotalSUBBvalue\ seconds. We calculated the SSAs for each initial rating in parallel on a CPU.
Therefore, we used Algorithm \ref{algo:ssa1} for all further calculations.


\begin{figure}
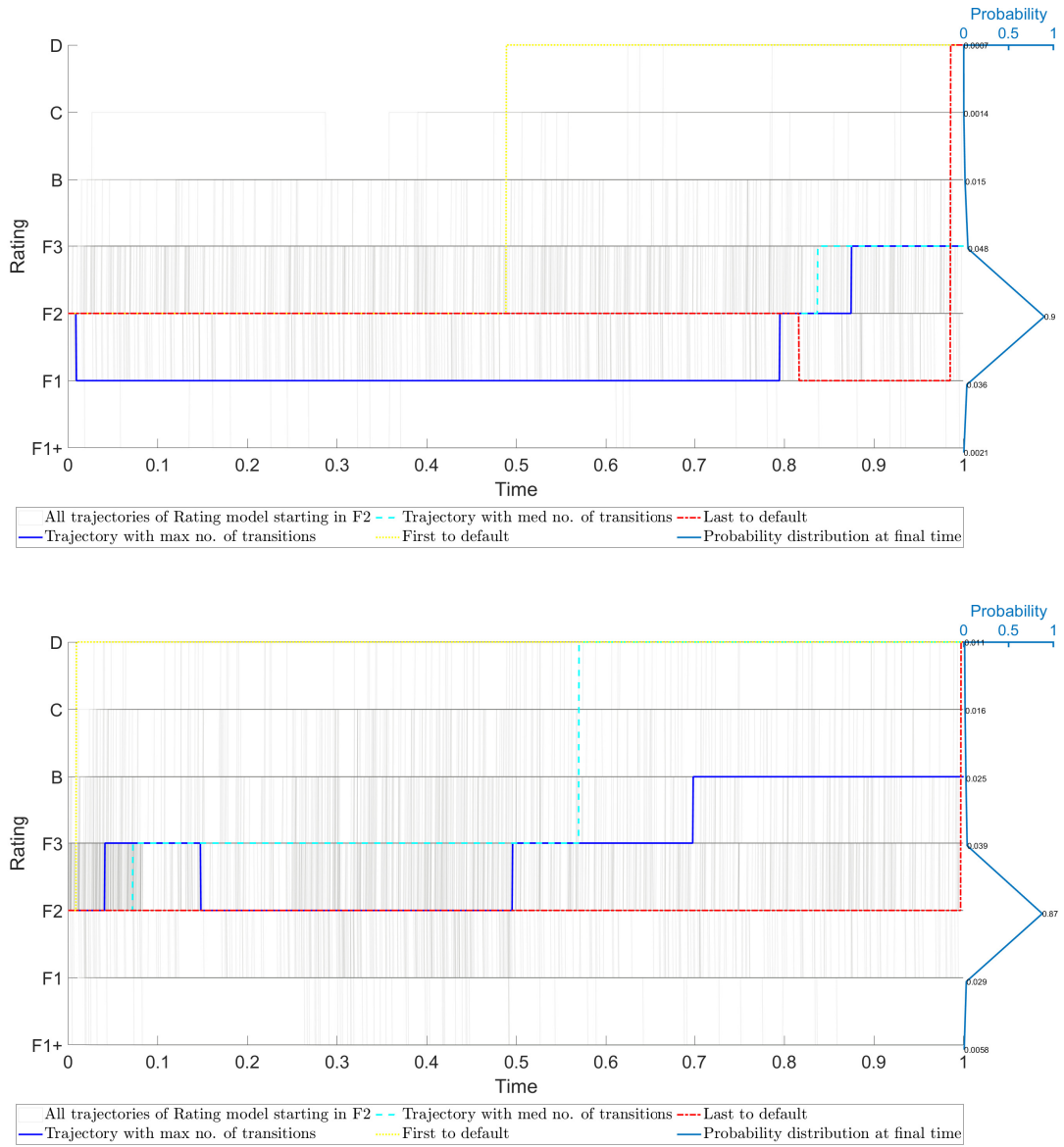
%
\figRatingPlotsPFOBBOne\\\noindent
\figRatingPlotsQFOBBOne
\caption{Simulated trajectories of PHCTMC $X_\cdot$ calibrated to Fitch's data set starting in rating F2. The top picture is under measure $\P$ and the bottom picture is under measure $\Q$.}%
\label{fig:RPFOBBOne}%
\end{figure}
\section{Application to rating triggers for collateral-inclusive bilateral valuation adjustments}\label{sec:ratingTriggers}
In this section, we discuss collateral-inclusive bilateral valuation adjustments with rating-dependent CSA thresholds. The main references for the general theory of XVA are 
\cite[\ pp.~375\,ff. Chapter 12.3 Credit Valuation Adjustment and Risk Management]{Oosterlee2019}
and
\cite[\ pp.~305\,ff. Chapter 13 Collateral, Netting, Close-Out and Re-Hypothecation]{Brigo2013}.
Throughout this section, we are taking the point of view of a bank having a portfolio of deals with a counterparty. The two parties have signed a netting set agreement with a CSA having rating-dependent thresholds. The stochastic process representing the future mark-to-market of the portfolio is denoted by $V_t$. We will assume that both contracting parties are subject to default and the default time will be denoted by $\tau_{B}$ and $\tau_{C}$ for the bank and the counterparty, respectively.
Additionally, we will suppose that the same rating matrices apply to both, i.e. they are in the same industrial sector and the consideration of two different sectors (cf. \cite{Bielecki2012} for a copula approach) is left for future research.
For our illustration, we will assume that the bank has the highest rating today, whereas the counterparty has a mid-range rating today, since we expect a bank to default less likely than the majority of companies. The evolution of their ratings over time will be denoted by $\ratingProcess_t^B$ and $\ratingProcess_t^C$, respectively, and we will set $\ratingProcess_t \coloneqq \left(\ratingProcess_t^B,\ratingProcess_t^C\right)$ to shorten notation.

Let $C_t$ be the stochastic process representing the value of the collateral account. In particular $C_t>0$ if the collateral is received by the bank. We will assume for simplicity that $C_t$ depends on $V$ only through the value $V_t$: more precisely we will suppose that $C_t \coloneqq f\left(V_t,\ratingProcess_t\right)$. In particular to avoid path dependencies we assume there are no minimum transfer amounts, the impact of this assumption being not material for our purposes.

We will discuss the following three scenarios of collateral agreements:


\begin{compactenum}
\item \emph{uncollateralized}, i.e. no collateral is interchanged and $f\equiv 0$;
\item \emph{perfectly collateralized}, i.e. collateral is posted instantaneously  at a discrete set of times, e.g. daily, and is equal to the mark-to-market ($V_t = C_t$), $f(v,r) = v$;
\item \emph{rating-trigger dependent}, more precisely we will focus on the case of thresholds depending on rating (see below for a description of $f$ in this case).
\end{compactenum}

The relation of the bank to the counterparty is illustrated in Figure \ref{fig:bankCpty}
and reads as follows:\footnote{
We will use the same conventions as in \cite[\ pp.~310\,ff. Chapter 13.2 Bilateral CVA Formula under Collateralization]{Brigo2013}, in particular $X^+=\max\left(X,0\right)$ and 
$X^-=\min\left(X,0\right)$.
}:

To illustrate the impact of collateral in risk mitigation, let us assume instantaneous posting and no rehypothecation of collateral for simplicity. Assume the counterparty (but not the bank) defaults at time $\tau$. Then on the one hand we have the value of the portfolio $V_\tau$ and on the other hand we have the value of the collateral account just before default $C_{\tau-}$. 
We distinguish four cases:
\begin{itemize}
\item $V_\tau\geq 0, C_{\tau^-}\geq 0$: the portfolio generates a positive exposure for the bank but this is mitigated by the collateral (which can be fully retrieved by the bank because of no rehypotecation). Therefore the outstanding claim is $V_t\tau - C_{\tau-}$. 
\item $V_\tau\geq 0, C_{\tau-} \leq 0$: although the portfolio generates a positive exposure for the bank, the bank had posted collateral just before default. Because of no rehypothecation, the bank can fully get back its collateral and the outstanding claim is therefore $V_\tau$.
\item  $V_\tau\leq 0, C_{\tau-} \geq 0$: the counterparty gets back the collateral posted to the bank and also gets the value of the portfolio $\abs{V_\tau}$.
\item $V_\tau\leq 0, C_{\tau-} \leq 0$: the counterparty keeps the collateral posted by the bank and also gets the remaining value of the portfolio $\abs{V_\tau-C_{\tau-}}$.
\end{itemize}
The behavior in case of default of the bank is symmetrical.

Additionally, the individual collateral postings depending on the collateral agreement are depicted by $f\left(V_t,\ratingProcess_t\right)^-$, meaning that the bank has to post collateral if this value is greater than zero and its analogue for the counterparty is given by $f\left(V_t,\ratingProcess_t\right)^+$.
 
A comprehensive explanation of all default events in this bilateral setup can be found in 
\cite[\ pp.~311--312 Chapter 13.2.1 Collection of CVA Contributions]{Brigo2013}.

\tikzset{mnode/.style={draw,rectangle,black,align=center,
											 text width = 2.5cm}}
\tikzset{marrow/.style={draw,thick}}
\begin{figure}%
\begin{minipage}[c][][c]{\linewidth}
\centering
\resizebox{\linewidth}{!}{
\begin{tikzpicture}
	\node[mnode] (bank) at (-7,0) {Bank};
	\node[mnode] (cpty) at (7,0) {Counterparty};
	\node[mnode] (portfolio) at (0,2) {Portfolio $V_t$};
	\node[mnode] (collateral) at (0,-3) {Collateral Account $C_t$};
	\node[mnode] (exposure) at (0,0) {Potential Loss $V_t-C_t$};
	\draw[->,marrow] (portfolio) -- (exposure);
	\draw[->,marrow] (collateral) -- (exposure);
	\draw[->,marrow] (exposure) -- node[midway,above] {$\left(V_t-C_t\right)^+$} (bank);
	\draw[->,marrow] (exposure) -- node[midway,above] {$\left(V_t-C_t\right)^-$} (cpty);
	
	\path[name path=bb] (bank.south west) -- (bank.south east);
	\path[name path=colb] ([yshift=2.5pt]collateral.west)  -- ([yshift=2.5pt]bank.south);
	\path[name intersections={of=bb and colb, by={ColB}}];
	\draw[-left to,marrow] 
		(ColB)
			-- node[text width= 3.75cm,midway,above,sloped,xshift=2.5em,align=center] {$f\left(V_t,\ratingProcess_t\right)^-<0$: post collateral} 
		([yshift=2.5pt] collateral.west) ;
	
	\path[name path=bcol] ([yshift=-2.5pt]collateral.west) -- ($([yshift=-2.5pt]collateral.west)!10cm!([yshift=-2.5pt]bank.south)$);
	\path [name intersections={of=bb and bcol, by={BCol}}];
	\draw[-left to,marrow] 
		([yshift=-2.5pt]collateral.west) 
			to node[midway,below,sloped] {$C_t>0$ in favor for} 
		(BCol);
	
	\path[name path=cc] (cpty.south west) -- (cpty.south east);
	\path[name path=cptyc] ([yshift=2.5pt]collateral.east)  -- ([yshift=2.5pt]cpty.south);
	\path[name intersections={of=cc and cptyc, by={CptyC}}];
	\draw[-right to,marrow] 
		(CptyC)
			-- node[text width= 3.75cm,midway,above,sloped,xshift=-2em,align=center] {$f\left(V_t,\ratingProcess_t\right)^+>0$: post collateral} 
		([yshift=2.5pt] collateral.east) ;
	
	\path[name path=ccpty] ([yshift=-2.5pt]collateral.east) -- ($([yshift=-2.5pt]collateral.east)!10cm!([yshift=-2.5pt]cpty.south)$);
	\path [name intersections={of=cc and ccpty, by={CCpty}}];
	\draw[-right to,marrow] 
		([yshift=-2.5pt]collateral.east) 
			to node[midway,below,sloped] {$C_t<0$ in favor for} 
		(CCpty);
	\draw[->,marrow] 
		(portfolio.west) 
			to node[midway,above,sloped] {$V_t>0$ creditor on default} 
		(bank.north);
	\draw[->,marrow] 
		(portfolio.east) 
			to node[midway,above,sloped] {$V_t<0$ creditor on default} 
		(cpty.north);
		
	\pgfresetboundingbox
	\path[use as bounding box] let \p{coordsBank} = (bank.west),
						\p{coordsCpty} = (cpty.east),
						\p{coordsC} = (collateral.south),
						\p{coordsV} = (portfolio.north)
				in
				($(\x{coordsBank},\y{coordsC})$) rectangle 
				($(\x{coordsCpty},\y{coordsV})$);
\end{tikzpicture}
}
\end{minipage}
\caption{Illustration of bank and counterparty relations in terms of exposure and collateral agreements.}%
\label{fig:bankCpty}%
\end{figure}
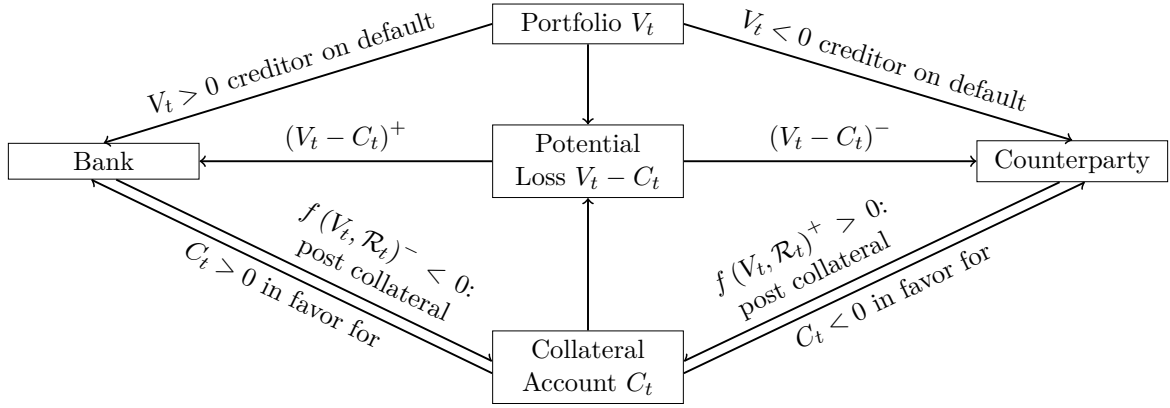

Next, we will discuss the impact of rating triggers compared to the aforementioned scenarios of collateral agreements on collateral-inclusive CVA, DVA and BVA, followed by a discussion on the pre-default distribution of the rating processes in 
Section \ref{sec:preDefault}.
\subsection{XVA with different collateral agreements}\label{sec:CXVA}
We are interested in the impact of rating triggers on \textCBVA, \textCCVA and \textCDVA (Collateralized Bilateral, Credit, Debit Valuation Adjustments) without the possibility of re-hypothecation and zero interest rate at mid-market to simplify the investigation.

Before we dive into this topic, let us first of all discuss our benchmark portfolio.
Since our main purpose is to analyse the general behavior of the rating model, we are not interested in setting up an accurate model for the computation of $V_t$. In particular the $V_t$ that we consider doesn't represent the value of a portfolio of real deals. Rather we decided to simulate $V_t$ using a number of independent Brownian motions with different volatilities and life-times to account for the cash-flows of the portfolio. To be more precise
\begin{align*}
	V_t \coloneqq 
	V_0+\sigma_0 W^0_t+\sum_{i=1}^{n}{
		\sigma_i W^i_t \1_{t\leq \tau^i}
	},
\end{align*}
where $V_0 \in \R_{\geq 0}$ is the initial value, $W^i$, $i=0,\dots,n$, $n\in\N$, are independent Brownian motions, $\sigma_i \in \R$ are volatilities and $\tau^i \in [0,T]$ are uniformly distributed random variables describing the different life-times of the cash-flows.
In the experiment we use $V_0=0$, $n=24$ and $\sigma_i$ are the standard normal random variables multiplied by $10$ for scaling and its sign indicates a positive or negative cash-flow (from the bank perspective). Also notice that we designed the portfolio in such a way that at least one cash-flow survives till $T$ by not adding a finite life-time to $W_t^0$. 
 
 Now, let us briefly recall the relevant definitions of \textCXVA (cf. \cite[\ p.~314 Equation 13.4]{Brigo2013}, \cite[\ p.~316 Equation 13.10]{Brigo2013}) without re-hypothecation
\begin{align}
	\CBVA\left(t,T,\collateral\right)&\coloneqq
	\CDVA\left(t,T,\collateral\right)-
	\CCVA\left(t,T,\collateral\right),
	\label{eq:CBVA}\\
	\CDVA\left(t,T,\collateral\right)&\coloneqq
	-\mathbb{E}^{\Q}\left[
		\left.
		\1_{\tau=\tau_B<T} \lgd_B \left(V_\tau^- - C_\tau^-\right)^-
		\right|
		\mathcal{G}_t
	\right],
	\label{eq:CDVA}\\
	\CCVA\left(t,T,\collateral\right)&\coloneqq
	\mathbb{E}^{\Q}\left[
		\left.
		\1_{\tau=\tau_C<T} \lgd_C \left(V_\tau^+ - C_\tau^+\right)^+
		\right|
		\mathcal{G}_t
	\right],
	\label{eq:CCVA}
\end{align}
where $\mathcal{G}_t$ is the filtration containing all the default-free market information plus default monitoring. These values are calculated under a risk-neutral measure $\Q$, which explains why we were interested in changing the measure of our rating model from the historical probabilities $\P$ to the risk-neutral measure in the first place.

The evaluation of the collateral account at the exact time of the default event, i.e. $C_\tau$, might seem confusing. We could imagine a scenario in which bonds or stocks could be used as collateral, making it necessary to evaluate the collateral account at the default event. In our case, we will assume that the collateral account will be a pure cash account, meaning that upon a default event the value will not be updated from its previous value $C_{\tau-}$. Therefore, it is very important to study the distribution of ratings prior to default, which is subject to Section \ref{sec:preDefault}.

We now describe the function $f\left(\exposure_{t},\ratingProcess_{t}\right)$ in the case of rating-triggers dependent agreements, following \cite[\ pp.~316\,ff. Chapter 13.5.2 Collateralization Through Margining]{Brigo2013}.
Let $r_i^{x}\geq 0$, $x\in \left\{B,C\right\}$, $i=1,\dots,K$ denote the threshold for the party $x$ in case $x$ has rating $i$: this means that the maximum unsecured exposure of the other party will be at most $r_i^{x}$. 

Now, we introduce the rating triggers $\rho^x$ with corresponding thresholds $r_i^x$ as
\begin{align*}
	\rho^x(i)\coloneqq
	\sum_{j=1}^{K}{r_j^x \1_{j}(i)}.
\end{align*}
As a small example, setting for all $i=1,\dots, K$ and $x=\left\{B,C\right\}$ the thresholds $r_i^x=+\infty$ leads to the uncollateralized scenario and $r_i^x=0$ to the perfectly collateralized scenario.

The amount of collateral to be posted by the bank at time $t_j$ is then
\begin{align*}
	\left(\exposure_{t_j} + \rho^B(\ratingProcess^B_{t_j})\right)^- -C_{t_j-}^-.
\end{align*}
For the counterparty we have analogously
\begin{align*}
	\left(\exposure_{t_j} - \rho^C(\ratingProcess^C_{t_j})\right)^+-C_{t_j-}^+.
\end{align*}

As aforementioned, we assume for simplicity that the value $C_{t_j}$ of the collateral account at time $t_j$ is equal to $C_{\beta(t_j)}$ where $\beta(u)$ is the last collateral posting date before $u$. In particular we assume there is no remuneration on the collateral account.
We then have
\begin{align*}
	C_{t_0}\coloneqq 0,\quad C_{t_n}\coloneqq 0, \quad C_{u-}\coloneqq
	C_{\beta(u)}.
\end{align*}
\begin{align*}
	C_{t_j}\coloneqq
	C_{t_j-}+
	\left(\left(\exposure_{t_j} + \rho^B(\ratingProcess^B_{t_j})\right)^- -C_{t_j-}^-\right)+
	\left(\left(\exposure_{t_j} - \rho^C(\ratingProcess^C_{t_j})\right)^+-C_{t_j-}^+\right).
\end{align*}
This can be rewritten as
\begin{align*}
	C_{t_j}=
		\left(\exposure_{t_j} + \rho^B(\ratingProcess^B_{t_j})\right)^-+
		\left(\exposure_{t_j} - \rho^C(\ratingProcess^C_{t_j})\right)^+
		\eqqcolon
		f\left(\exposure_{t_j},\ratingProcess_{t_j}\right).
\end{align*}

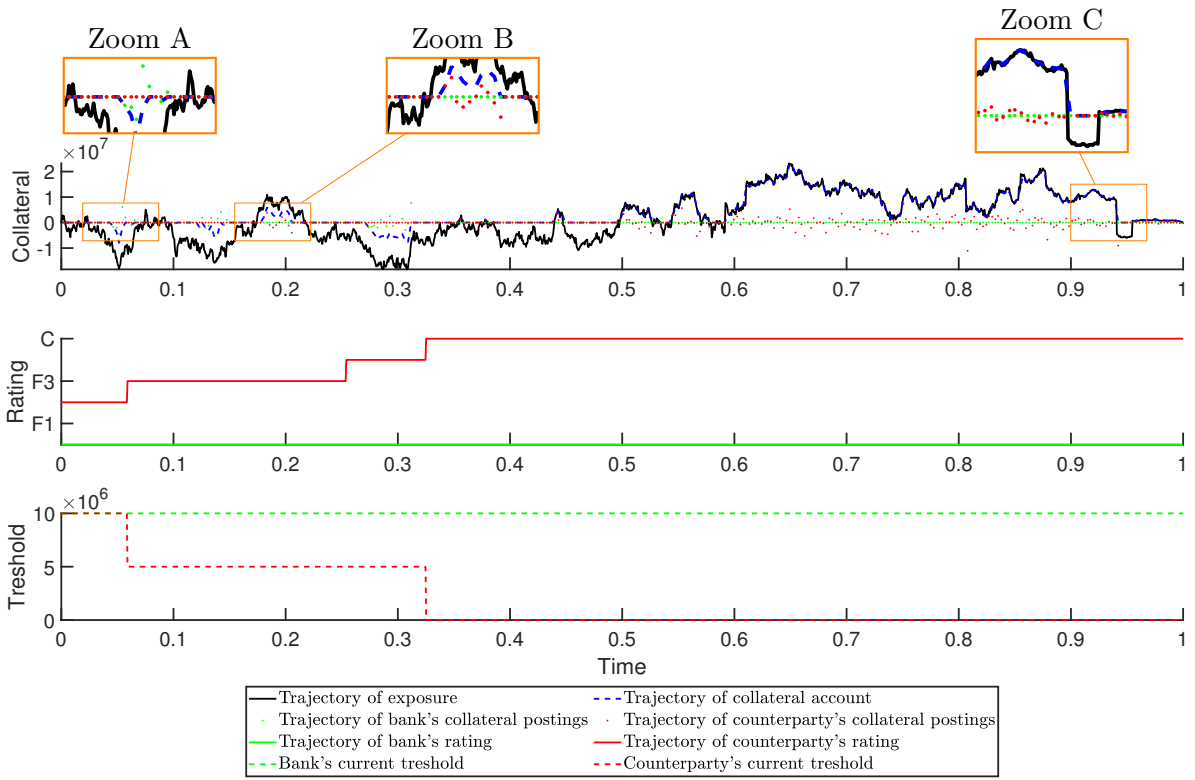
\begin{figure}%
\begin{tikzpicture}[spy using outlines={rectangle, connect spies}]
	\node[draw=none] at (0,0) {\figCollateralRTFOBBTwo};
	
	\spy [orange, draw, height = 1cm, width = 2cm, magnification = 2] on (-6.25,3.125) in node at (-6,4.8);
	\node[anchor=south,align=center] at (-6,5.3) {Zoom A};
	
	\spy [orange, draw, height = 1cm, width = 2cm, magnification = 2] on (-4.25,3.125) in node at (-1.75,4.8);
	\node[anchor=south,align=center] at (-1.75,5.3) {Zoom B};
	
	\spy [orange, draw, height = 1.5cm, width = 2cm, magnification = 2] on (6.75,3.25) in node at (6,4.8);
	\node[anchor=south,align=center] at (6,5.55) {Zoom C};
\end{tikzpicture}
\caption{One trajectory of a collateral agreement with rating triggers using Fitch's data. Top picture shows the collateral account and portfolio over time, middle one the rating evolution and bottom one the corresponding rating thresholds.}%
\label{fig:collateralFOBBTwo}%
\end{figure}

We will use in all experiments 365 posting dates per year.
In Figure \ref{fig:collateralFOBBTwo} 
one can see one trajectory of the portfolio, collateral account and individual postings by both counterparties in the top picture. The picture in the middle indicates the ratings of both counterparties over time for this particular trajectory and the bottom picture shows the corresponding threshold for each point in time. The orange boxes are magnifications of the indicated sections in the figure.

One can see that for this choice of trajectory the bank has no rating transition and the counterparty has many, ranging through all thresholds defined in Table \ref{tab:ThresholdsFOBB}.

At the section ``Zoom A'' one can see that only the bank has to post collateral (green dots), since the portfolio is negative and the allowed threshold is exceeded. In section ``Zoom B'' we can see the first impact due to the change of the countparty's threshold, which now has to post collateral (red dots). We can see that the distance between the blue dashed line (the collateral account) and the black line (the portfolio) is closer than e.g. in ``Zoom A'', because the threshold is now five million euros. After time $t=0.35$ the rating of the counterparty reaches \textbf{C}. Therefore, the threshold is set to zero and we can see that the collateral stays close to the value of the portfolio, whenever its positive, which is also the content of ``Zoom C''.

\begin{table}%
\caption{Rating thresholds for both counterparties using Fitch's data. Values in Euro.}
\centering
\begin{tabular}{|*{7}{c}|}
	\hline
	F1+ & F1 & F2 & F3 & B & C & D \\
	\hline
	\multicolumn{7}{|c|}{Bank}\\
	$1e+07$ & $1e+07$ & $1e+07$ & $5e+06$ & $5e+06$ & $ 0$ & $ 0$ \\
	\hline
	\multicolumn{7}{|c|}{Counterparty}\\
	$1e+07$ & $1e+07$ & $1e+07$ & $5e+06$ & $5e+06$ & $ 0$ & $ 0$\\
	\hline
\end{tabular}
\label{tab:ThresholdsFOBB}
\end{table}


In Table \ref{tab:CXVAFOBB}
are the values of \textCDVA \eqref{eq:CDVA}, \textCCVA \eqref{eq:CCVA} and \textCBVA \eqref{eq:CBVA} using
the Loss-Given-Default
$\lgd_B=0.6$, $\lgd_C=0.6$ and $M=10000$ simulations for 
the three collateral agreements: no collateralization, perfect collateralization and collateralization with rating triggers.

One can see, that the collateralization with rating triggers lies in between the values of the uncollateralized case and the perfectly collateralized case, which is the expected behaviour, because as illustrated in Figure \ref{fig:collateralFOBBTwo} one has a possible transition from unsecured money to the perfectly collaterlized scenario, where rating thresholds are zero. The difference to the perfectly collateralized case is that there can be transitions from high ratings to default in one instant, which will be subject of the next subsection.
\begin{table}%
\centering
\caption{\textCXVA with the different collateral agreements (no, perfectly and rating triggers) using Fitch's data and $\lgd_B=0.6$, as well as $\lgd_C=0.6$ 
 with  $M=10000$ simulations and thresholds defined in 
 Table \ref{tab:ThresholdsFOBB}.}
\begin{tabular}{|l|*{3}{c}|}
	\hline
	CXVA		& Uncollateralized 	 & Rating Triggers 		& Perfectly collateralized\\
	\hline
	CDVA 		& $9.13e+ 04$ & $5.05e + 04$ & $2.39e + 04$\\
	CCVA 		& $9.96e + 04$ & $5.67e + 04$ & $3.09e + 04$\\
	CBVA 		& $-8.3e + 03$ & $-6.23e + 03$ & $-7.07e + 03$\\
	\hline
\end{tabular}
\label{tab:CXVAFOBB}
\end{table}

\subsection{Pre-default rating distribution}\label{sec:preDefault}
\begin{figure}[ht]
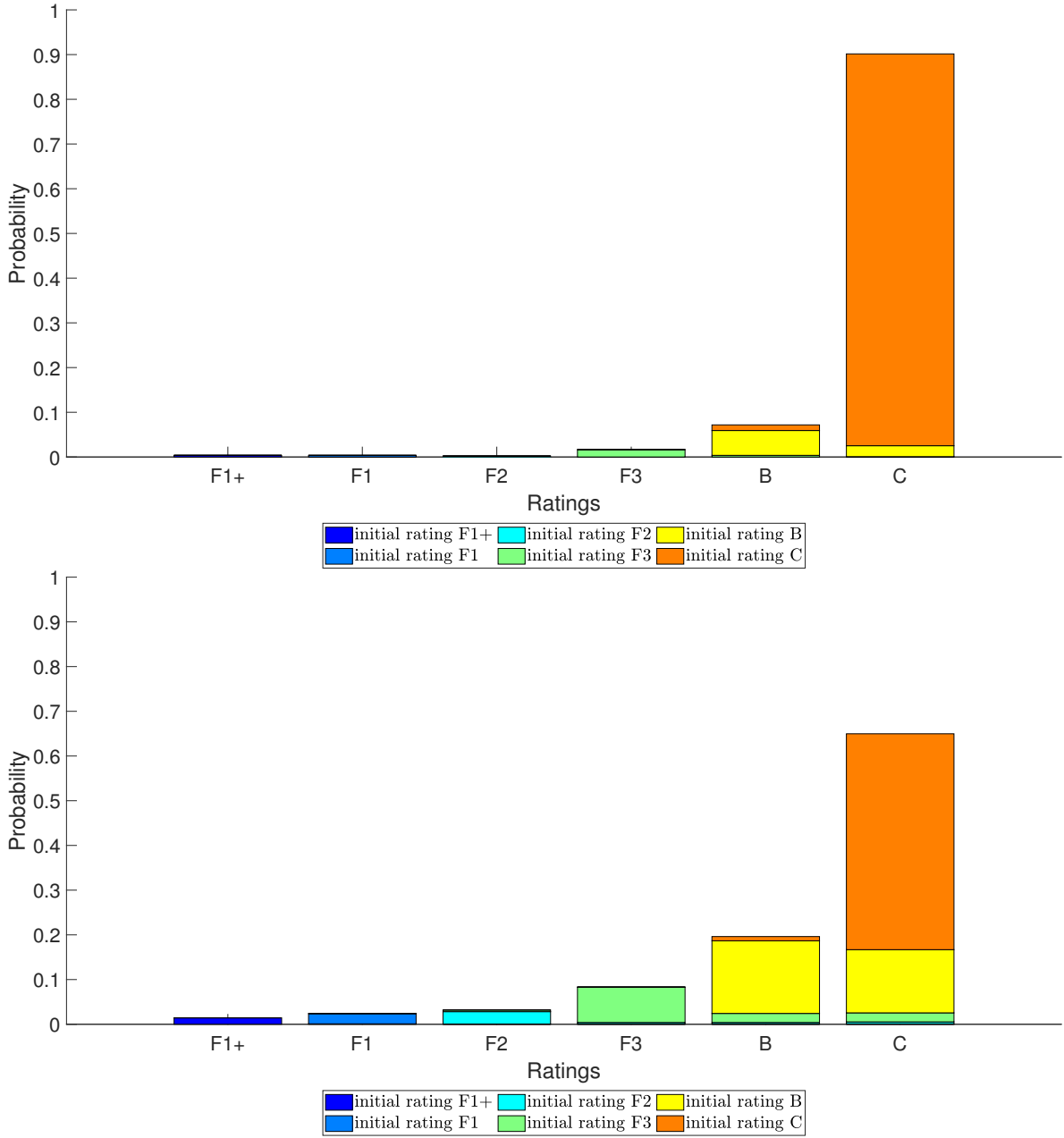
%
\figPrePDPFOBBSeven\\\noindent
\figPrePDQFOBBSeven
\caption{Pre-default distribution using Fitch's data. The upper picture is under the historical measure and the lower plot under the risk-neutral measure.}%
\label{fig:PrePDFOBBSeven}%
\end{figure}

As it is apparent from the definition of the rating thresholds, it is important to study the distribution of the rating process one time-instant prior to default, because this will determine the unsecured amount of money at the default event. We will call this henceforth pre-default distribution and will also compare the distribution under $\P$ to the one under $\Q$ with the help of Figure \ref{fig:PrePDFOBBSeven}, which were obtained by Monte-Carlo simulation.

Now, let us have a closer look at Figure \ref{fig:PrePDFOBBSeven}. 
First of all, one can see the pre-default distribution
under the measure $\P$ in the top picture and under the measure $\Q$ in the bottom picture.
Disregarding the individual colors, the probability of being in a certain rating prior to default is given by the total height of the column. 
The composition of the individual colors of each column indicate the contribution of each starting rating, e.g. in the column prior to default we can see that the most prominent contributions are resulting from the two ratings prior to default, e.g. in Fitch's data rating B and C, but there are contributions of the other ratings as well.

In the market, it can be observed that the default probabilities in the risk neutral world are usually higher than the default probabilities quoted under the historical measure in the rating matrices. This phenomenon has an impact in our model on all other ratings as well, which can already be seen in Figure \ref{fig:RPFOBBOne} 
by the spread of the grey lines indicating all simulated trajectories. In the risk-neutral world there seem to be more transitions than in the historical world causing this spread of grey lines. 
The reason for this is that the calibration of this model has essentially one parameter for each rating because $h_i\in \R^K$ for each time interval where the process is homogeneous. Therefore, the high probability of default under the measure $\Q$ compared to the one under $\P$ has a significant impact on the other ratings as well.

We can see that under the measure $\P$, the top picture, almost all the defaults had a prior rating of C, while under the measure $\Q$, this is still the most prominent case but significantly smaller. It is more likely under the measure $\Q$ that a company starting with a high rating defaults and this without transitioning to the rating prior to default first, which is indicated by the different heights of the each individual color for each rating.

It is yet an open question and needs thorough economical investigation whether this behavior makes sense or not, because it has a significant impact on the performance of collateralization with rating triggers. To be more precise, the more likely it is, that a company starting in a good rating defaults without first transitioning to a rating, where a low threshold is defined, the more unsecured money we have at a default event.
\section{Conclusion and future research}\label{sec:conclusion}
In this paper, we have seen how to use piecewise homogeneous continuous-time Markov chains to model rating transitions. We have demonstrated how the most popular change of measure techniques perform on our dataset and found an improvement by introducing a penalized calibration procedure. After that, we computed \textCCVA and
\textCDVA by simulating rating processes from the rating transitions by means of the stochastic simulation algorithm. We showed that rating triggers lowered both
\textCCVA and \textCDVA as expected and discussed the pre-default distribution in the end.

Allowing for small errors of the generator in the calibration under the historical measure revealed a high sensitivity of the model to its generator. In this line of thought, the impact of the initial adjustment of the defect rating matrices published by the agencies might be very severe. In future research, we would like to consider a stochastic model for rating transitions, maybe including a stochastic rating matrix adjustment to reduce the sensitivity of the model. A first step towards this direction has been taken in \cite{KM2022} by means of Machine Learning and SDEs on Lie-Groups.

Moreover, for the application to rating triggers it will be important that the entities can be in different sectors. Including the possibility of copulae as in \cite{Bielecki2012} will be left for future research as well.
\appendix
\section{Model data}\label{sec:modelData}
In this appendix we list the rating transitions under the historical measure and under the risk-neutral measure after calibration with $M^{\P}\equiv 1 \equiv m^{\Q}$.
\begin{table}[ht]%
\centering
\caption{Rating matrix under measure $\P$ for one month using Fitch's data.}
\begin{tabular}{|c|*{7}{c}|}
\hline
\diagbox{From}{To} & F1+ & F1 & F2 & F3 & B & C & D\\ \hline
F1+ & $99.366$\,\% & $0.613$\,\% & $0.020$\,\% & $0.000$\,\% & $0.000$\,\% & $0.000$\,\% & $0.000$\,\% \\
F1 & $0.261$\,\% & $99.086$\,\% & $0.622$\,\% & $0.030$\,\% & $0.000$\,\% & $0.000$\,\% & $0.000$\,\% \\
F2 & $0.030$\,\% & $0.281$\,\% & $99.085$\,\% & $0.512$\,\% & $0.090$\,\% & $0.000$\,\% & $0.000$\,\% \\
F3 & $0.010$\,\% & $0.040$\,\% & $0.725$\,\% & $98.388$\,\% & $0.806$\,\% & $0.020$\,\% & $0.010$\,\% \\
B & $0.000$\,\% & $0.000$\,\% & $0.040$\,\% & $0.343$\,\% & $99.223$\,\% & $0.363$\,\% & $0.030$\,\% \\
C & $0.000$\,\% & $0.000$\,\% & $0.001$\,\% & $0.005$\,\% & $2.886$\,\% & $95.201$\,\% & $1.907$\,\% \\
D & $0.000$\,\% & $0.000$\,\% & $0.000$\,\% & $0.000$\,\% & $0.000$\,\% & $0.000$\,\% & $100.000$\,\% 
\\\hline
\end{tabular}
\label{tab:ratingMatrixAnalyticAPFOBBOne}
\end{table}
\begin{table}[ht]%
\centering
\caption{Rating matrix under measure $\P$ for three months using Fitch's data.}
\begin{tabular}{|c|*{7}{c}|}
\hline
\diagbox{From}{To} & F1+ & F1 & F2 & F3 & B & C & D\\ \hline
F1+ & $98.079$\,\% & $1.818$\,\% & $0.081$\,\% & $0.020$\,\% & $0.001$\,\% & $0.000$\,\% & $0.000$\,\% \\
F1 & $0.758$\,\% & $97.300$\,\% & $1.810$\,\% & $0.091$\,\% & $0.030$\,\% & $0.000$\,\% & $0.010$\,\% \\
F2 & $0.091$\,\% & $0.842$\,\% & $97.241$\,\% & $1.491$\,\% & $0.304$\,\% & $0.010$\,\% & $0.020$\,\% \\
F3 & $0.041$\,\% & $0.123$\,\% & $2.219$\,\% & $95.225$\,\% & $2.331$\,\% & $0.041$\,\% & $0.020$\,\% \\
B & $0.000$\,\% & $0.010$\,\% & $0.133$\,\% & $1.057$\,\% & $97.630$\,\% & $1.025$\,\% & $0.144$\,\% \\
C & $0.000$\,\% & $0.001$\,\% & $0.006$\,\% & $0.049$\,\% & $8.991$\,\% & $85.931$\,\% & $5.023$\,\% \\
D & $0.000$\,\% & $0.000$\,\% & $0.000$\,\% & $0.000$\,\% & $0.000$\,\% & $0.000$\,\% & $100.000$\,\% 
\\\hline
\end{tabular}
\label{tab:ratingMatrixAnalyticAPFOBBTwo}
\end{table}
\begin{table}[ht]%
\centering
\caption{Rating matrix under measure $\P$ for six months using Fitch's data.}
\begin{tabular}{|c|*{7}{c}|}
\hline
\diagbox{From}{To} & F1+ & F1 & F2 & F3 & B & C & D\\ \hline
F1+ & $96.151$\,\% & $3.570$\,\% & $0.206$\,\% & $0.041$\,\% & $0.010$\,\% & $0.000$\,\% & $0.021$\,\% \\
F1 & $1.503$\,\% & $94.682$\,\% & $3.486$\,\% & $0.215$\,\% & $0.092$\,\% & $0.002$\,\% & $0.020$\,\% \\
F2 & $0.165$\,\% & $1.699$\,\% & $94.615$\,\% & $2.811$\,\% & $0.638$\,\% & $0.041$\,\% & $0.031$\,\% \\
F3 & $0.084$\,\% & $0.230$\,\% & $4.406$\,\% & $90.801$\,\% & $4.333$\,\% & $0.084$\,\% & $0.063$\,\% \\
B & $0.001$\,\% & $0.021$\,\% & $0.264$\,\% & $2.174$\,\% & $95.310$\,\% & $1.829$\,\% & $0.400$\,\% \\
C & $0.000$\,\% & $0.002$\,\% & $0.023$\,\% & $0.216$\,\% & $18.610$\,\% & $72.787$\,\% & $8.362$\,\% \\
D & $0.000$\,\% & $0.000$\,\% & $0.000$\,\% & $0.000$\,\% & $0.000$\,\% & $0.000$\,\% & $100.000$\,\% 
\\\hline
\end{tabular}
\label{tab:ratingMatrixAnalyticAPFOBBThree}
\end{table}
\begin{table}[ht]%
\centering
\caption{Rating matrix under measure $\P$ for twelve months using Fitch's data.}
\begin{tabular}{|c|*{7}{c}|}
\hline
\diagbox{From}{To} & F1+ & F1 & F2 & F3 & B & C & D\\ \hline
F1+ & $92.470$\,\% & $6.753$\,\% & $0.585$\,\% & $0.096$\,\% & $0.043$\,\% & $0.001$\,\% & $0.053$\,\% \\
F1 & $2.845$\,\% & $89.782$\,\% & $6.485$\,\% & $0.523$\,\% & $0.305$\,\% & $0.008$\,\% & $0.053$\,\% \\
F2 & $0.276$\,\% & $3.380$\,\% & $89.732$\,\% & $4.954$\,\% & $1.456$\,\% & $0.105$\,\% & $0.096$\,\% \\
F3 & $0.209$\,\% & $0.439$\,\% & $9.059$\,\% & $82.653$\,\% & $7.181$\,\% & $0.206$\,\% & $0.252$\,\% \\
B & $0.007$\,\% & $0.034$\,\% & $0.513$\,\% & $4.522$\,\% & $90.826$\,\% & $3.046$\,\% & $1.053$\,\% \\
C & $0.003$\,\% & $0.004$\,\% & $0.076$\,\% & $0.989$\,\% & $37.477$\,\% & $49.160$\,\% & $12.292$\,\% \\
D & $0.000$\,\% & $0.000$\,\% & $0.000$\,\% & $0.000$\,\% & $0.000$\,\% & $0.000$\,\% & $100.000$\,\% 
\\\hline
\end{tabular}
\label{tab:ratingMatrixAnalyticAPFOBBFour}
\end{table}
\begin{table}[ht]%
\centering
\caption{Rating matrix under measure $\Q$ for one month using Fitch's data.}
\begin{tabular}{|c|*{7}{c}|}
\hline
\diagbox{From}{To} & F1+ & F1 & F2 & F3 & B & C & D\\ \hline
F1+ & $99.469$\,\% & $0.418$\,\% & $0.010$\,\% & $0.002$\,\% & $0.002$\,\% & $0.059$\,\% & $0.042$\,\% \\
F1 & $0.383$\,\% & $98.942$\,\% & $0.453$\,\% & $0.070$\,\% & $0.003$\,\% & $0.086$\,\% & $0.062$\,\% \\
F2 & $0.060$\,\% & $0.379$\,\% & $97.394$\,\% & $1.617$\,\% & $0.314$\,\% & $0.144$\,\% & $0.092$\,\% \\
F3 & $0.006$\,\% & $0.017$\,\% & $0.224$\,\% & $97.662$\,\% & $0.879$\,\% & $0.897$\,\% & $0.314$\,\% \\
B & $0.000$\,\% & $0.000$\,\% & $0.010$\,\% & $0.271$\,\% & $86.058$\,\% & $12.906$\,\% & $0.754$\,\% \\
C & $0.000$\,\% & $0.000$\,\% & $0.000$\,\% & $0.000$\,\% & $0.073$\,\% & $98.549$\,\% & $1.378$\,\% \\
D & $0.000$\,\% & $0.000$\,\% & $0.000$\,\% & $0.000$\,\% & $0.000$\,\% & $0.000$\,\% & $100.000$\,\% 
\\\hline
\end{tabular}
\label{tab:ratingMatrixAnalyticAQFOBBOne}
\end{table}
\begin{table}[ht]%
\centering
\caption{Rating matrix under measure $\Q$ for three months using Fitch's data.}
\begin{tabular}{|c|*{7}{c}|}
\hline
\diagbox{From}{To} & F1+ & F1 & F2 & F3 & B & C & D\\ \hline
F1+ & $94.201$\,\% & $5.182$\,\% & $0.317$\,\% & $0.012$\,\% & $0.008$\,\% & $0.153$\,\% & $0.127$\,\% \\
F1 & $0.485$\,\% & $96.977$\,\% & $2.146$\,\% & $0.063$\,\% & $0.032$\,\% & $0.112$\,\% & $0.186$\,\% \\
F2 & $0.068$\,\% & $0.779$\,\% & $96.702$\,\% & $1.361$\,\% & $0.511$\,\% & $0.300$\,\% & $0.280$\,\% \\
F3 & $0.048$\,\% & $0.461$\,\% & $12.141$\,\% & $76.121$\,\% & $7.866$\,\% & $2.425$\,\% & $0.939$\,\% \\
B & $0.001$\,\% & $0.011$\,\% & $0.159$\,\% & $0.307$\,\% & $74.814$\,\% & $22.464$\,\% & $2.245$\,\% \\
C & $0.000$\,\% & $0.000$\,\% & $0.001$\,\% & $0.000$\,\% & $0.431$\,\% & $95.492$\,\% & $4.077$\,\% \\
D & $0.000$\,\% & $0.000$\,\% & $0.000$\,\% & $0.000$\,\% & $0.000$\,\% & $0.000$\,\% & $100.000$\,\% 
\\\hline
\end{tabular}
\label{tab:ratingMatrixAnalyticAQFOBBTwo}
\end{table}
\begin{table}[ht]%
\centering
\caption{Rating matrix under measure $\Q$ for six months using Fitch's data.}
\begin{tabular}{|c|*{7}{c}|}
\hline
\diagbox{From}{To} & F1+ & F1 & F2 & F3 & B & C & D\\ \hline
F1+ & $93.630$\,\% & $5.530$\,\% & $0.374$\,\% & $0.024$\,\% & $0.025$\,\% & $0.164$\,\% & $0.253$\,\% \\
F1 & $2.868$\,\% & $93.168$\,\% & $3.081$\,\% & $0.162$\,\% & $0.173$\,\% & $0.178$\,\% & $0.371$\,\% \\
F2 & $0.427$\,\% & $2.115$\,\% & $91.390$\,\% & $2.759$\,\% & $1.575$\,\% & $1.176$\,\% & $0.559$\,\% \\
F3 & $0.241$\,\% & $0.710$\,\% & $12.960$\,\% & $68.795$\,\% & $11.911$\,\% & $3.514$\,\% & $1.870$\,\% \\
B & $0.005$\,\% & $0.016$\,\% & $0.178$\,\% & $0.553$\,\% & $67.670$\,\% & $27.138$\,\% & $4.440$\,\% \\
C & $0.001$\,\% & $0.000$\,\% & $0.001$\,\% & $0.005$\,\% & $1.788$\,\% & $90.218$\,\% & $7.987$\,\% \\
D & $0.000$\,\% & $0.000$\,\% & $0.000$\,\% & $0.000$\,\% & $0.000$\,\% & $0.000$\,\% & $100.000$\,\% 
\\\hline
\end{tabular}
\label{tab:ratingMatrixAnalyticAQFOBBThree}
\end{table}
\begin{table}[ht]%
\centering
\caption{Rating matrix under measure $\Q$ for twelve months using Fitch's data.}
\begin{tabular}{|c|*{7}{c}|}
\hline
\diagbox{From}{To} & F1+ & F1 & F2 & F3 & B & C & D\\ \hline
F1+ & $91.273$\,\% & $6.951$\,\% & $0.938$\,\% & $0.073$\,\% & $0.092$\,\% & $0.169$\,\% & $0.505$\,\% \\
F1 & $5.070$\,\% & $84.704$\,\% & $8.207$\,\% & $0.458$\,\% & $0.582$\,\% & $0.238$\,\% & $0.741$\,\% \\
F2 & $0.558$\,\% & $2.910$\,\% & $87.492$\,\% & $3.919$\,\% & $2.533$\,\% & $1.474$\,\% & $1.115$\,\% \\
F3 & $0.389$\,\% & $0.871$\,\% & $17.517$\,\% & $58.447$\,\% & $14.750$\,\% & $4.322$\,\% & $3.704$\,\% \\
B & $0.010$\,\% & $0.019$\,\% & $0.290$\,\% & $1.390$\,\% & $61.375$\,\% & $28.233$\,\% & $8.682$\,\% \\
C & $0.002$\,\% & $0.001$\,\% & $0.008$\,\% & $0.081$\,\% & $8.393$\,\% & $76.179$\,\% & $15.336$\,\% \\
D & $0.000$\,\% & $0.000$\,\% & $0.000$\,\% & $0.000$\,\% & $0.000$\,\% & $0.000$\,\% & $100.000$\,\% 
\\\hline
\end{tabular}
\label{tab:ratingMatrixAnalyticAQFOBBFour}
\end{table}
\section{Historical and market data}\label{sec:marketData}
In this appendix  we listed the rating matrices used in this paper. They contain the transition probabilities at a certain point for a particular period of time. For example in Table \ref{tab:ratingMatrixMarketFOBBOne} the probability to stay at rating \verb=F1+=, if one started in \verb=F1+=, over the period of one month is $98.84\,\%$.
\begin{table}[ht]%
\centering
\caption{Rating matrix under measure $\P$ for one month using Fitch's data.}
\begin{tabular}{|c|*{7}{c}|}
\hline
\diagbox{From}{To} & F1+ & F1 & F2 & F3 & B & C & D\\ \hline
F1+ & $98.840$\,\% & $0.610$\,\% & $0.020$\,\% & $0.000$\,\% & $0.000$\,\% & $0.000$\,\% & $0.000$\,\% \\
F1 & $0.260$\,\% & $98.740$\,\% & $0.620$\,\% & $0.030$\,\% & $0.000$\,\% & $0.000$\,\% & $0.000$\,\% \\
F2 & $0.030$\,\% & $0.280$\,\% & $98.610$\,\% & $0.510$\,\% & $0.090$\,\% & $0.000$\,\% & $0.000$\,\% \\
F3 & $0.010$\,\% & $0.040$\,\% & $0.720$\,\% & $97.680$\,\% & $0.800$\,\% & $0.020$\,\% & $0.010$\,\% \\
B & $0.000$\,\% & $0.000$\,\% & $0.040$\,\% & $0.340$\,\% & $98.380$\,\% & $0.360$\,\% & $0.030$\,\% \\
C & $0.000$\,\% & $0.000$\,\% & $0.000$\,\% & $0.000$\,\% & $2.830$\,\% & $93.350$\,\% & $1.870$\,\% \\
D & $0.000$\,\% & $0.000$\,\% & $0.000$\,\% & $0.000$\,\% & $0.000$\,\% & $0.000$\,\% & $100.000$\,\% 
\\\hline
\end{tabular}
\label{tab:ratingMatrixMarketFOBBOne}
\end{table}
\begin{table}[ht]%
\centering
\caption{Rating matrix under measure $\P$ for three months using Fitch's data.}
\begin{tabular}{|c|*{7}{c}|}
\hline
\diagbox{From}{To} & F1+ & F1 & F2 & F3 & B & C & D\\ \hline
F1+ & $96.550$\,\% & $1.790$\,\% & $0.080$\,\% & $0.020$\,\% & $0.000$\,\% & $0.000$\,\% & $0.000$\,\% \\
F1 & $0.750$\,\% & $96.230$\,\% & $1.790$\,\% & $0.090$\,\% & $0.030$\,\% & $0.000$\,\% & $0.010$\,\% \\
F2 & $0.090$\,\% & $0.830$\,\% & $95.870$\,\% & $1.470$\,\% & $0.300$\,\% & $0.010$\,\% & $0.020$\,\% \\
F3 & $0.040$\,\% & $0.120$\,\% & $2.170$\,\% & $93.140$\,\% & $2.280$\,\% & $0.040$\,\% & $0.020$\,\% \\
B & $0.000$\,\% & $0.010$\,\% & $0.130$\,\% & $1.030$\,\% & $95.170$\,\% & $1.000$\,\% & $0.140$\,\% \\
C & $0.000$\,\% & $0.000$\,\% & $0.000$\,\% & $0.000$\,\% & $8.520$\,\% & $81.470$\,\% & $4.760$\,\% \\
D & $0.000$\,\% & $0.000$\,\% & $0.000$\,\% & $0.000$\,\% & $0.000$\,\% & $0.000$\,\% & $100.000$\,\% 
\\\hline
\end{tabular}
\label{tab:ratingMatrixMarketFOBBTwo}
\end{table}
\begin{table}[ht]%
\centering
\caption{Rating matrix under measure $\P$ for six months using Fitch's data.}
\begin{tabular}{|c|*{7}{c}|}
\hline
\diagbox{From}{To} & F1+ & F1 & F2 & F3 & B & C & D\\ \hline
F1+ & $93.190$\,\% & $3.460$\,\% & $0.200$\,\% & $0.040$\,\% & $0.010$\,\% & $0.000$\,\% & $0.020$\,\% \\
F1 & $1.470$\,\% & $92.610$\,\% & $3.410$\,\% & $0.210$\,\% & $0.090$\,\% & $0.000$\,\% & $0.020$\,\% \\
F2 & $0.160$\,\% & $1.650$\,\% & $91.900$\,\% & $2.730$\,\% & $0.620$\,\% & $0.040$\,\% & $0.030$\,\% \\
F3 & $0.080$\,\% & $0.220$\,\% & $4.210$\,\% & $86.760$\,\% & $4.140$\,\% & $0.080$\,\% & $0.060$\,\% \\
B & $0.000$\,\% & $0.020$\,\% & $0.250$\,\% & $2.060$\,\% & $90.460$\,\% & $1.740$\,\% & $0.380$\,\% \\
C & $0.000$\,\% & $0.000$\,\% & $0.000$\,\% & $0.000$\,\% & $16.870$\,\% & $66.140$\,\% & $7.580$\,\% \\
D & $0.000$\,\% & $0.000$\,\% & $0.000$\,\% & $0.000$\,\% & $0.000$\,\% & $0.000$\,\% & $100.000$\,\% 
\\\hline
\end{tabular}
\label{tab:ratingMatrixMarketFOBBThree}
\end{table}
\begin{table}[ht]%
\centering
\caption{Rating matrix under measure $\P$ for twelve months using Fitch's data.}
\begin{tabular}{|c|*{7}{c}|}
\hline
\diagbox{From}{To} & F1+ & F1 & F2 & F3 & B & C & D\\ \hline
F1+ & $86.950$\,\% & $6.350$\,\% & $0.550$\,\% & $0.090$\,\% & $0.040$\,\% & $0.000$\,\% & $0.050$\,\% \\
F1 & $2.720$\,\% & $85.850$\,\% & $6.200$\,\% & $0.500$\,\% & $0.290$\,\% & $0.000$\,\% & $0.050$\,\% \\
F2 & $0.260$\,\% & $3.180$\,\% & $84.420$\,\% & $4.660$\,\% & $1.370$\,\% & $0.100$\,\% & $0.090$\,\% \\
F3 & $0.190$\,\% & $0.400$\,\% & $8.250$\,\% & $75.270$\,\% & $6.540$\,\% & $0.190$\,\% & $0.230$\,\% \\
B & $0.000$\,\% & $0.030$\,\% & $0.460$\,\% & $4.040$\,\% & $81.890$\,\% & $2.780$\,\% & $0.950$\,\% \\
C & $0.000$\,\% & $0.000$\,\% & $0.000$\,\% & $0.000$\,\% & $31.880$\,\% & $42.400$\,\% & $10.440$\,\% \\
D & $0.000$\,\% & $0.000$\,\% & $0.000$\,\% & $0.000$\,\% & $0.000$\,\% & $0.000$\,\% & $100.000$\,\% 
\\\hline
\end{tabular}
\label{tab:ratingMatrixMarketFOBBFour}
\end{table}
\begin{table}%
\centering
\caption{Default probabilities under measure $\Q$ for $1,3,6,12$ months using Fitch's data.}
\begin{tabular}{|c|*{4}{c}|}
\hline
\diagbox{From}{To} & D $\left(t=\frac{1}{12}\right)$ & D $\left(t=\frac{3}{12}\right)$ & D $\left(t=\frac{6}{12}\right)$ & D $\left(t=\frac{12}{12}\right)$ \\
\hline
F1+ & $0.042$\,\% & $0.127$\,\% & $0.253$\,\% & $0.505$\,\% \\
F1 & $0.062$\,\% & $0.186$\,\% & $0.371$\,\% & $0.741$\,\% \\
F2 & $0.093$\,\% & $0.280$\,\% & $0.559$\,\% & $1.115$\,\% \\
F3 & $0.314$\,\% & $0.939$\,\% & $1.870$\,\% & $3.704$\,\% \\
B & $0.754$\,\% & $2.245$\,\% & $4.440$\,\% & $8.682$\,\% \\
C & $1.378$\,\% & $4.077$\,\% & $7.987$\,\% & $15.336$\,\% \\
D & $100.000$\,\% & $100.000$\,\% & $100.000$\,\% & $100.000$\,\% \\
\hline
\end{tabular}%
\label{tab:defaultProbabilitiesFOBB}
\end{table}

\section*{Declarations}
\subsection*{Funding}
This project has received funding from the European Union’s Horizon 2020 research and innovation
programme under the Marie Sklodowska-Curie grant agreement No 813261 and is part of the ABC-EU-XVA project.
\subsection*{Conflicts of interests}

The authors have no relevant financial or non-financial interests to disclose.

\subsection*{Data availability}
All data generated or analysed during this study are included in this published article.
{
In particular the code to produce the numerical experiments is available at\\
\url{https://github.com/kevinkamm/RatingTriggers}.
}
{\thispagestyle{scrheadings}
\newpage
\thispagestyle{scrheadings}\ihead{}
\singlespacing
\AtNextBibliography{\small}
\printbibliography
}
\end{document}